\documentclass[namedate,webpdf,modern,medium]{oup-authoring-template}
\onecolumn
\usepackage{microtype}
\microtypesetup{nopatch=footnote}
\usepackage{mathtools,booktabs,tabularx,longtable,pdflscape,needspace,setspace,placeins}

\setcitestyle{authoryear,round,semicolon}
\hypersetup{hidelinks}
\newcommand{\E}{\mathbb{E}}
\newcommand{\Var}{\operatorname{Var}}

\newcommand{\expit}{\operatorname{expit}}

\newcommand{\G}{\mathbb{G}}

\newcommand{\societylogo}{}

\makeatletter
\def\ps@opening{\let\@oddhead\@empty\let\@evenhead\@empty
  \def\@oddfoot{\hfil\thepage\hfil}\let\@evenfoot\@oddfoot}
\def\ps@headings{
  \def\@oddhead{\small\itshape Estimand-specific bootstrap confidence intervals\hfil}
  \def\@evenhead{\small\itshape Takano et al.\hfil}
  \def\@oddfoot{\hfil\thepage\hfil}\let\@evenfoot\@oddfoot}
\makeatother
\begin{document}
\journaltitle{}
\DOI{}\copyrightyear{}\pubyear{}\vol{}\issue{}\access{}\appnotes{}
\firstpage{1}
\title[Estimand-specific bootstrap confidence intervals]{Estimand-specific bootstrap confidence intervals after optimism correction for prediction models}
\author[1,2,$\ast$]{Seiya Takano\ORCID{0000-0001-6990-9923}}
\author[3,4]{Kazuki Nishida}
\author[4]{Ryo Emoto}
\author[1,4]{Shigeyuki Matsui}
\address[1]{Department of Biostatistics and Data Science, School of Public Health, Graduate School of Medicine, Kyoto University, Yoshida-Konoe-cho, Sakyo-ku, Kyoto 606-8501, Japan}
\address[2]{Department of Radiology, Nagoya City University Graduate School of Medical Sciences, 1 Kawasumi, Mizuho-cho, Mizuho-ku, Nagoya, Aichi 467-8601, Japan}
\address[3]{Department of Biostatistics, Nagoya University Graduate School of Medicine, 65 Tsurumai-cho, Showa-ku, Nagoya, Aichi 466-8550, Japan}
\address[4]{Department of Clinical Biostatistics, School of Public Health, Graduate School of Medicine, Kyoto University, Yoshida-Konoe-cho, Sakyo-ku, Kyoto 606-8501, Japan}
\corresp[$\ast$]{Corresponding author: Seiya Takano, M.D., Ph.D., Department of Biostatistics and Data Science, School of Public Health, Graduate School of Medicine, Kyoto University, Yoshida-Konoe-cho, Sakyo-ku, Kyoto 606-8501, Japan. Phone: (+81) 75-753-4482; E-mail: \href{mailto:stakano6213@gmail.com}{stakano6213@gmail.com}}

\abstract{Confidence intervals after bootstrap optimism correction are often reported for prediction models, but what they cover depends on the estimand. We distinguish two estimands: the target-population performance of the final model developed from the observed sample, $\theta_{\mathrm{FM},n}$, and the average performance over repeated model development, $\theta_{\mathrm{proc},n}$. Using first-order expansions of the estimation errors, we identify the bootstrap root that matches the error for $\theta_{\mathrm{FM},n}$. We use this root in the nested error bootstrap (NEB) as a new reference method, and we show that the optimism error bootstrap (OEB), which needs no inner bootstrap, approximates NEB. We also show that the conventional two-stage bootstrap (TSB) and location-shifted bootstrap (LSB) reproduce the first-order variation for $\theta_{\mathrm{proc},n}$. In simulations, NEB generally achieved coverage closer to the nominal level than TSB and LSB for the area under the receiver operating characteristic curve, Brier score, and log loss. OEB was close to NEB at moderate to high events per variable (EPV) but could give narrower intervals at low EPV. The distinction is particularly clear for calibration measures: first-order effects of perturbing the training and evaluation distributions cancel in the TSB root, yielding a degenerate first-order limit. In contrast, the NEB root has a nondegenerate first-order limit and yields asymptotically valid confidence intervals for $\theta_{\mathrm{FM},n}$ under regularity conditions. Consistent with theory, TSB showed substantial undercoverage for calibration even at larger sample sizes. Our findings emphasize the need to construct confidence intervals for a specified estimand and performance measure.}
\begingroup\setlength{\parindent}{0pt}\maketitle\endgroup

\section{Introduction}\label{sec:introduction}

When a prediction model is evaluated in its development sample, its performance on new observations from the target population is generally overestimated. This overestimation is referred to as optimism \citep{efron1983}. In clinical prediction, bootstrap internal validation is widely recommended for estimating optimism-corrected prediction performance \citep{harrell1996,steyerberg2001}. Harrell's bias correction is an established method for point estimation \citep{harrell1996}, whereas methods for constructing confidence intervals (CIs) after optimism correction are less standardized. Noma et al. proposed the two-stage bootstrap (TSB), which recalculates the corrected statistic within each bootstrap sample, and the location-shifted bootstrap (LSB), which shifts the apparent bootstrap distribution and is computationally simpler \citep{noma2021}.

A remaining issue is that the uncertainty represented by a CI depends on the estimand. For a prediction model developed from the observed sample, two distinct performance estimands can be considered. The first is the target-population performance of the final model developed from that sample, denoted by $\theta_{\mathrm{FM},n}$. The second is the average performance over hypothetical repetitions of  the model-development procedure, denoted by $\theta_{\mathrm{proc},n}$. The distinction between these two estimands is well recognized in the literature on prediction error and cross-validation \citep{efron1983,bates2024}. Estimation errors for these two estimands can involve different components of sampling variation. An optimism-corrected point estimate alone therefore does not determine the appropriate CI \citep{noma2021}. The bootstrap distribution used to construct the CI should approximate the sampling distribution of estimation error for the estimand being reported \citep{bickelFreedman1981,vanderVaart1998}. However, whether existing bootstrap CI methods after optimism correction reproduce the relevant variation has not been fully characterized.

The aim of this paper is to develop bootstrap CIs for $\theta_{\mathrm{FM},n}$ after optimism correction. We focus on $\theta_{\mathrm{FM},n}$ because bootstrap optimism correction is commonly used to assess how the model developed from the observed sample will perform on new observations from the target population \citep{harrell1996,steyerberg2001,noma2021,collinsEvaluation2024}. To guide the construction of CIs, we derive first-order asymptotic expansions of the estimation errors for the two estimands. We then examine whether the distribution of each bootstrap root approximates that of the leading term in the corresponding expansion. This framework leads to two main results. First, we identify the bootstrap root that matches the estimation error for $\theta_{\mathrm{FM},n}$. We implement it in the nested error bootstrap (NEB), a reference method, and derive the optimism error bootstrap (OEB) as a one-level approximation that fixes Harrell's bias correction across outer bootstrap samples. One-level replicates of the same form as OEB have been used before, and our results give them a justification \citep{baart2019,meisner2020,martin2021}. Second, we show that TSB and LSB reproduce the first-order variation for $\theta_{\mathrm{proc},n}$ rather than the estimation error for $\theta_{\mathrm{FM},n}$. For calibration slope, calibration intercept, and calibration-in-the-large (CITL), the distinction between the two estimands has a particularly strong consequence. Under unpenalized logistic regression fitted by maximum likelihood, the first-order variation represented by TSB and LSB is degenerate on the $\sqrt n$ scale, whereas NEB retains the nonzero first-order variation required for inference on $\theta_{\mathrm{FM},n}$. This contrast illustrates why the choice of bootstrap distribution depends on both the estimand and the performance measure. We also evaluate the performance of the proposed methods in simulation studies and illustrate their application using the GUSTO-I West region data.

\section{Methods}\label{sec:methods}

\subsection{Performance functional and estimands}\label{sec:estimands}

We consider a prediction model developed from one sample and evaluated on new observations from a target population. Let $Z_i=(X_i,Y_i)$, $i=1,\ldots,n$, be independent observations from $P_0$, where $X_i$ is the predictor vector and $Y_i$ is the outcome, and let $P_n$ be the empirical distribution of the observed development sample. For a probability distribution $P$ and a $P$-integrable function $f$, write $Pf=\int f(z)\,dP(z)$. In particular, $P_n f=n^{-1}\sum_{i=1}^{n}f(Z_i)$. A prespecified learning procedure applied to a training distribution $P$ returns a prediction rule $\eta_P$. We write $\Psi(P,Q)$ for the performance of this rule under an evaluation distribution $Q$. For smooth mean-type measures, including Brier score and log loss, $\Psi(P,Q)=\E_Q\{m_P(Z)\}$, where $m_P(z)=\ell\{y,\eta_P(x)\}$ and $\ell$ is the loss function for the chosen measure.

The primary estimand is the performance on new observations from the target population of the final model developed using the observed sample, 
\begin{equation}
\theta_{\mathrm{FM},n}=\Psi(P_n,P_0).
\label{eq:theta-fm}
\end{equation}
For a probability distribution $P$, let $P^n$ denote the joint distribution of $n$ independent observations from $P$, and let $P_n^\dagger$ denote the empirical distribution of an independent hypothetical development sample of size $n$ from $P_0$. An alternative estimand is the average performance over hypothetical repetitions of the model-development procedure,
\begin{equation}
\theta_{\mathrm{proc},n}
=
\E_{P_0^n}\{\Psi(P_n^\dagger,P_0)\}.
\label{eq:theta-proc}
\end{equation}
Under the first-order differentiability conditions in Supporting Information Section~S1, let $\phi_{\mathrm{tr}}$ and $\phi_{\mathrm{ev}}$ denote the influence functions for perturbations of the training and evaluation distributions, respectively, and let $\G_n f=\sqrt n(P_n-P_0)f$. For any corrected point estimator $C_n(P_n)$ satisfying
\[
\sqrt n\{C_n(P_n)-\Psi(P_n,P_n)\}=o_p(1),
\]
the corresponding estimation errors satisfy
\begin{equation}
\begin{aligned}
\sqrt n\{C_n(P_n)-\theta_{\mathrm{FM},n}\}
&=\G_n\phi_{\mathrm{ev}}+o_p(1),\\
\sqrt n\{C_n(P_n)-\theta_{\mathrm{proc},n}\}
&=\G_n(\phi_{\mathrm{tr}}+\phi_{\mathrm{ev}})+o_p(1).
\end{aligned}
\label{eq:generic-estimand-roots}
\end{equation}
The first-order variation in the error for $\theta_{\mathrm{FM},n}$ is given by $\phi_{\mathrm{ev}}$ alone because the estimator and target contain the same fitted rule. The error for $\theta_{\mathrm{proc},n}$ instead has first-order variation given by $\phi_{\mathrm{tr}}+\phi_{\mathrm{ev}}$ because its target averages over a new development sample. Accordingly, we distinguish the bootstrap CI methods by the first-order variation represented by their bootstrap roots. A formal derivation is given in Supporting Information Section~S1. Additional conditions for AUC are given in Supporting Information Section~S3. The calibration measures are considered separately in Section~\ref{sec:calibration-structure} and Appendix~\ref{app:calibration}.

\subsection{Point estimation and bootstrap roots for the two estimands}\label{sec:estimators-roots}

For any probability distribution $P$, let $P_n^\star$ and $P_n^\dagger$ denote empirical distributions of samples of $n$ independent observations from $P$. The bootstrap estimate of optimism is
\begin{equation}
\Omega_{H,n}(P)
=
\E_{P^n}
\left\{
\Psi(P_n^\star,P_n^\star)
-
\Psi(P_n^\star,P)
\right\}.
\label{eq:bootstrap-optimism}
\end{equation}
Using this quantity, define
\begin{equation}
\begin{aligned}
C_{H,n}(P)
&=
\Psi(P,P)-\Omega_{H,n}(P),\\
C_{\mathrm{proc},n}(P)
&=
\E_{P^n}
\{\Psi(P_n^\dagger,P)\}.
\end{aligned}
\label{eq:point-estimators}
\end{equation}
When $P=P_n$, $C_{H,n}(P_n)$ and $C_{\mathrm{proc},n}(P_n)$ are the point estimators considered for $\theta_{\mathrm{FM},n}$ and $\theta_{\mathrm{proc},n}$, respectively. Specifically, $C_{H,n}(P_n)$ is the usual estimator obtained by Harrell's bias correction \citep{harrell1996,efron1983}, whereas $C_{\mathrm{proc},n}(P_n)$ is the nonparametric plug-in estimator for $\theta_{\mathrm{proc},n}$ because $C_{\mathrm{proc},n}(P_0)=\theta_{\mathrm{proc},n}$.

Accordingly, the estimation-error roots for $\theta_{\mathrm{FM},n}$ and $\theta_{\mathrm{proc},n}$ are
\[
\sqrt n\{C_{H,n}(P_n)-\theta_{\mathrm{FM},n}\}
\qquad\text{and}\qquad
\sqrt n\{C_{\mathrm{proc},n}(P_n)-\theta_{\mathrm{proc},n}\},
\]
respectively. Their first-order expansions are given in Equation~\eqref{eq:generic-estimand-roots}. The empirical bootstrap replaces the population distribution $P_0$ by $P_n$ and the observed empirical distribution $P_n$ by $P_n^\star$. Applying this substitution gives
\begin{equation}
\begin{aligned}
R_{\mathrm{FM},n}^\star
&=
\sqrt n\left\{
C_{H,n}(P_n^\star)
-
\Psi(P_n^\star,P_n)
\right\},\\
R_{\mathrm{proc},n}^\star
&=
\sqrt n\left\{
C_{\mathrm{proc},n}(P_n^\star)
-
C_{\mathrm{proc},n}(P_n)
\right\}.
\end{aligned}
\label{eq:reference-roots}
\end{equation}
Under the conditions in Supporting Information Section~S1, the conditional distributions of these bootstrap roots approximate the sampling distributions of the corresponding estimation-error roots. We use them as references when defining and comparing the bootstrap methods below.

\subsection{Bootstrap confidence intervals}\label{sec:algorithms}

Let $P_n^{\star b}$ denote the empirical distribution of the $b$th outer bootstrap sample, $b=1,\ldots,B_{\mathrm{outer}}$, and let $P_n^{\star\star(b,j)}$, $j=1,\ldots,B_{\mathrm{inner}}$, denote the empirical distribution of the $j$th inner bootstrap sample drawn from $P_n^{\star b}$ when nested resampling is required. Write
\[
A_n=\Psi(P_n,P_n),
\qquad
A_b^\star=\Psi(P_n^{\star b},P_n^{\star b}).
\]
For method $\mathcal M$, let $D_b^{\mathcal M}$ denote the $b$th bootstrap replicate used to form the CI. The $1-\alpha$ CI is
\begin{equation}
CI_{1-\alpha}^{\mathcal M}
=
\left[
q_{\alpha/2}\{D_b^{\mathcal M}\},
q_{1-\alpha/2}\{D_b^{\mathcal M}\}
\right],
\label{eq:generic-ci}
\end{equation}
where $q_u$ denotes the empirical $u$ quantile over the outer bootstrap replicates.
The Apparent interval uses $D_b^{\mathrm{Apparent}}=A_b^\star$ in Equation~\eqref{eq:generic-ci}; no optimism correction is applied.

\subsubsection{Reference and approximation for \texorpdfstring{$\theta_{\mathrm{FM},n}$}{theta FM}: NEB and OEB}\label{sec:neb-oeb}

For $\theta_{\mathrm{FM},n}$, Equation~\eqref{eq:reference-roots} gives the bootstrap analog of the estimation error $C_{H,n}(P_n)-\theta_{\mathrm{FM},n}$. Replacing this unknown error by its bootstrap analog in $\theta_{\mathrm{FM},n}=C_{H,n}(P_n)-\{C_{H,n}(P_n)-\theta_{\mathrm{FM},n}\}$ gives the nested error bootstrap (NEB),
\begin{equation}
D_b^{\mathrm{NEB}}
=
C_{H,n}(P_n)
-
\left\{
C_{H,n}(P_n^{\star b})
-
\Psi(P_n^{\star b},P_n)
\right\}.
\label{eq:neb-pseudovalue}
\end{equation}
Computing $C_{H,n}(P_n^{\star b})$ requires estimating the optimism correction within each outer bootstrap sample and therefore requires an inner bootstrap. Algorithm~\ref{alg:neb} gives the Monte Carlo implementation.

\begin{algorithm}[H]
\caption{Nested error bootstrap for $\theta_{\mathrm{FM},n}$}
\label{alg:neb}
\begin{algorithmic}[1]
\Statex \textbf{Input:} $P_n$, $B_{\mathrm{outer}}$, $B_{\mathrm{inner}}$, $\alpha$
\State Fit the prespecified learning procedure to $P_n$ and compute the apparent performance $A_n=\Psi(P_n,P_n)$.

\For{$k=1,\ldots,B_{\mathrm{outer}}$}
    \State Draw $\widetilde P_n^{\star k}$ from $P_n$ and refit the learning procedure.
    \State Compute $\widetilde A_k^\star=\Psi(\widetilde P_n^{\star k},\widetilde P_n^{\star k})$ and $\Psi(\widetilde P_n^{\star k},P_n)$.
\EndFor
\State Set
\[
\widehat C_{H,n}(P_n)
=
A_n-
\frac{1}{B_{\mathrm{outer}}}
\sum_{k=1}^{B_{\mathrm{outer}}}
\left\{
\widetilde A_k^\star-\Psi(\widetilde P_n^{\star k},P_n)
\right\}.
\]

\For{$b=1,\ldots,B_{\mathrm{outer}}$}
    \State Independently of the bootstrap samples above, draw $P_n^{\star b}$ from $P_n$ and refit the learning procedure.
    \State Compute $A_b^\star=\Psi(P_n^{\star b},P_n^{\star b})$ and $\Psi(P_n^{\star b},P_n)$.
    \For{$j=1,\ldots,B_{\mathrm{inner}}$}
        \State Draw $P_n^{\star\star(b,j)}$ from $P_n^{\star b}$ and refit the learning procedure.
        \State Compute $A_{bj}^{\star\star}=\Psi(P_n^{\star\star(b,j)},P_n^{\star\star(b,j)})$ and $\Psi(P_n^{\star\star(b,j)},P_n^{\star b})$.
    \EndFor
    \State Set
    \[
    \widehat\Omega_{H,n}(P_n^{\star b})
    =
    \frac{1}{B_{\mathrm{inner}}}
    \sum_{j=1}^{B_{\mathrm{inner}}}
    \left\{
    A_{bj}^{\star\star}
    -
    \Psi(P_n^{\star\star(b,j)},P_n^{\star b})
    \right\},
    \qquad
    \widehat C_{H,n}(P_n^{\star b})
    =
    A_b^\star-\widehat\Omega_{H,n}(P_n^{\star b}).
    \]
    \State Form
    \[
    D_b
    =
    \widehat C_{H,n}(P_n)
    -
    \left\{
    \widehat C_{H,n}(P_n^{\star b})
    -
    \Psi(P_n^{\star b},P_n)
    \right\}.
    \]
\EndFor

\Statex \textbf{Output:} The empirical $\alpha/2$ and $1-\alpha/2$ quantiles of $\{D_b\}_{b=1}^{B_{\mathrm{outer}}}$.
\end{algorithmic}
\end{algorithm}

OEB is a one-level approximation to NEB that avoids the inner bootstrap by fixing the optimism correction at its observed-sample value across outer bootstrap samples. Specifically,
\[
C_{H,n}(P_n^{\star b})
=
A_b^\star-\Omega_{H,n}(P_n^{\star b})
\approx
A_b^\star-\Omega_{H,n}(P_n).
\]
Substituting this approximation into Equation~\eqref{eq:neb-pseudovalue} gives the optimism error bootstrap (OEB),
\begin{equation}
D_b^{\mathrm{OEB}}
=
A_n-A_b^\star+\Psi(P_n^{\star b},P_n).
\label{eq:oeb-pseudovalue}
\end{equation}
Let $\Delta\Omega_{H,b}=\Omega_{H,n}(P_n^{\star b})-\Omega_{H,n}(P_n).$ Then the exact paired difference is
\begin{equation}
D_b^{\mathrm{NEB}}
=
D_b^{\mathrm{OEB}}
+
\Delta\Omega_{H,b}.
\label{eq:neb-oeb-difference}
\end{equation}
Consequently, if
\begin{equation}
\sqrt n\,\Delta\Omega_{H,b}=o_{p^\star}(1),
\label{eq:oeb-first-order-condition}
\end{equation}
OEB and NEB have the same first-order limiting distribution. The formal transfer result is given in Supporting Information Section~S2.2. Algorithm~\ref{alg:oeb} gives the Monte Carlo implementation.

\begin{algorithm}[H]
\caption{Optimism error bootstrap for $\theta_{\mathrm{FM},n}$}
\label{alg:oeb}
\begin{algorithmic}[1]
\Statex \textbf{Input:} $P_n$, $B_{\mathrm{outer}}$, $\alpha$
\State Fit the prespecified learning procedure to $P_n$ and compute the apparent performance $A_n=\Psi(P_n,P_n)$.

\For{$b=1,\ldots,B_{\mathrm{outer}}$}
    \State Draw $P_n^{\star b}$ from $P_n$ and refit the learning procedure.
    \State Compute $A_b^\star=\Psi(P_n^{\star b},P_n^{\star b})$ and $\Psi(P_n^{\star b},P_n)$.
    \State Set
    \[
    O_b^\star
    =
    A_b^\star-\Psi(P_n^{\star b},P_n).
    \]
\EndFor

\State Set
\[
\widehat\Omega_{H,n}(P_n)
=
\frac{1}{B_{\mathrm{outer}}}
\sum_{b=1}^{B_{\mathrm{outer}}} O_b^\star,
\qquad
\widehat C_{H,n}(P_n)
=
A_n-\widehat\Omega_{H,n}(P_n).
\]

\State For $b=1,\ldots,B_{\mathrm{outer}}$, form
\[
D_b
=
\widehat C_{H,n}(P_n)
-
\left\{
O_b^\star-\widehat\Omega_{H,n}(P_n)
\right\}
=
A_n-A_b^\star+\Psi(P_n^{\star b},P_n).
\]

\Statex \textbf{Output:} The empirical $\alpha/2$ and $1-\alpha/2$ quantiles of $\{D_b\}_{b=1}^{B_{\mathrm{outer}}}$.
\end{algorithmic}
\end{algorithm}

Bootstrap replicates algebraically identical to those of OEB have been used previously \citep{baart2019,meisner2020,martin2021}, although their relation to $\theta_{\mathrm{FM},n}$ and to a nested reference method was not established. Harrell proposed asymmetric bootstrap confidence limits for overfitting-corrected model performance metrics (ABCLOC) in an online report \citep{harrell2025}. Like OEB, ABCLOC uses the performance of each bootstrap-fitted model in the bootstrap and original samples. Under the conditions in Supporting Information Section~S6.2, ABCLOC with $c=1$ has the same first-order limit as OEB, and for the calibration measures, $c>1$ gives asymptotic coverage above the nominal level. The derivation and simulation comparisons are given in Supporting Information Section~S6.

\subsubsection{Reference and approximation for \texorpdfstring{$\theta_{\mathrm{proc},n}$}{theta proc}: bTSB and bLSB}

For a point estimate $\widehat\theta$ and its bootstrap replicate $\widehat\theta^\star$, the basic bootstrap method uses $\widehat\theta-(\widehat\theta^\star-\widehat\theta)=2\widehat\theta-\widehat\theta^\star$, thereby reversing the bootstrap deviation from the observed estimate \citep{davisonHinkley1997}.
For $\theta_{\mathrm{proc},n}$, applying this method to $C_{\mathrm{proc},n}$ gives the basic two-stage bootstrap (bTSB):
\begin{equation}
D_b^{\mathrm{bTSB}}
=
2C_{\mathrm{proc},n}(P_n)-C_{\mathrm{proc},n}(P_n^{\star b}).
\label{eq:btsb-pseudovalue}
\end{equation}
We refer to its one-level approximation as the basic location-shifted bootstrap (bLSB):
\begin{equation}
D_b^{\mathrm{bLSB}}
=
C_{\mathrm{proc},n}(P_n)+A_n-A_b^\star.
\label{eq:blsb-pseudovalue}
\end{equation}
Let $\Delta\Omega_{\mathrm{proc},b}=\left\{A_b^\star-C_{\mathrm{proc},n}(P_n^{\star b})\right\}-\left\{A_n-C_{\mathrm{proc},n}(P_n)
\right\}.$ Then
\[
D_b^{\mathrm{bTSB}}-D_b^{\mathrm{bLSB}}
=
\Delta\Omega_{\mathrm{proc},b}.
\]
Therefore, if
\begin{equation}
\sqrt n\,\Delta\Omega_{\mathrm{proc},b}
=
o_{p^\star}(1),
\label{eq:blsb-first-order-condition}
\end{equation}
bTSB and bLSB have the same first-order limiting distribution.

\subsubsection{TSB and LSB}

Noma et al. proposed TSB and LSB \citep{noma2021}. As shown in Section~\ref{sec:first-order-roots}, both methods retain the first-order variation in the estimation error for $\theta_{\mathrm{proc},n}$. Their bootstrap replicates are
\begin{equation}
\begin{aligned}
D_b^{\mathrm{TSB}}
&=
C_{H,n}(P_n^{\star b}),\\
D_b^{\mathrm{LSB}}
&=
C_{H,n}(P_n)+A_b^\star-A_n.
\end{aligned}
\label{eq:historical-values}
\end{equation}
For TSB and LSB, the quantiles in Equation~\eqref{eq:generic-ci} are taken directly from the bootstrap replicates in Equation~\eqref{eq:historical-values}, as in the percentile bootstrap method \citep{davisonHinkley1997}. Using the same $\Delta\Omega_{H,b}$,
\[
D_b^{\mathrm{TSB}}-D_b^{\mathrm{LSB}}
=
-\Delta\Omega_{H,b}.
\]
Thus, under Equation~\eqref{eq:oeb-first-order-condition}, TSB and LSB have the same first-order limiting distribution. For the calibration measures, as shown below in Equation~\eqref{eq:calibration-diagonal}, TSB and bTSB use the same corrected estimate in each outer bootstrap sample and therefore have the same interval width when the conditional inner-bootstrap expectation is used, although the two intervals can have different lower and upper confidence limits.

\subsection{First-order expansions for the bootstrap methods}\label{sec:first-order-roots}

Let $\G_n^\star=\sqrt n(P_n^{\star b}-P_n)$ conditionally on the observed sample. Under Equations~\eqref{eq:oeb-first-order-condition} and~\eqref{eq:blsb-first-order-condition}, together with the conditions in Supporting Information Section~S1,
\begin{equation}
\begin{aligned}
\sqrt n\{D_b^{\mathcal M}-C_{H,n}(P_n)\}
&=-\G_n^\star\phi_{\mathrm{ev}}+o_{p^\star}(1),
&&\mathcal M\in\{\mathrm{NEB},\mathrm{OEB}\},\\
\sqrt n\{D_b^{\mathcal M}-C_{H,n}(P_n)\}
&=+\G_n^\star(\phi_{\mathrm{tr}}+\phi_{\mathrm{ev}})+o_{p^\star}(1),
&&\mathcal M\in\{\mathrm{TSB},\mathrm{LSB}\},\\
\sqrt n\{D_b^{\mathcal M}-C_{\mathrm{proc},n}(P_n)\}
&=-\G_n^\star(\phi_{\mathrm{tr}}+\phi_{\mathrm{ev}})+o_{p^\star}(1),
&&\mathcal M\in\{\mathrm{bTSB},\mathrm{bLSB}\}.
\end{aligned}
\label{eq:method-root-map}
\end{equation}
NEB and OEB therefore retain the first-order variation for $\theta_{\mathrm{FM},n}$, whereas TSB, LSB, bTSB, and bLSB retain that for $\theta_{\mathrm{proc},n}$. In Equation~\eqref{eq:method-root-map}, the leading centered term for TSB and LSB has the opposite sign from that for bTSB and bLSB. Because the limiting Gaussian distribution is centered and symmetric, this sign change does not alter the first-order limiting distribution, although the resulting intervals can differ in finite samples.

The first-order component retained by a bootstrap root does not by itself establish bootstrap validity. NEB and bTSB arise directly from the bootstrap analogs of the estimation errors for their respective estimands, whereas OEB and bLSB additionally require the corresponding change in the correction to be negligible on the $\sqrt n$ scale. The distinction can disappear for particular performance measures: under unpenalized logistic regression fitted by maximum likelihood, $\phi_{\mathrm{tr}}=0$ for log loss, and the same cancellation holds for Brier score under correct logistic specification. Additional conditions for AUC with a fitted prediction score are given in Supporting Information Section~S3. Formal derivations are given in Supporting Information Sections~S1--S3.

\subsection{First-order behavior of confidence intervals for calibration measures}\label{sec:calibration-structure}

For a logistic model fitted under $P$, let $\eta_P(X)$ denote its linear predictor. Under $Q$, the calibration intercept $\alpha_{\mathrm{cal}}(P,Q)$ and slope $\gamma_{\mathrm{cal}}(P,Q)$ are obtained jointly from logistic regression of $Y$ on $\eta_P(X)$. We define CITL, denoted by $\alpha_{\mathrm{CITL}}(P,Q)$, as the intercept from the same regression with the slope fixed at 1 \citep{vanCalster2016,riley2021}. We write $\Psi_{\mathrm{cal}}(P,Q)$ for any one of these three measures. Appendix~\ref{app:calibration} gives their estimating equations and formal derivations of Equations~\eqref{eq:calibration-diagonal}--\eqref{eq:calibration-root-separation} below.

For these calibration measures, the distinction between the two estimands is particularly clear. When logistic regression with an intercept is fitted by maximum likelihood without penalization and evaluated under the same distribution, the score equations give
\begin{equation}
\Psi_{\mathrm{cal}}(P,P)
=
\begin{cases}
1, & \text{for calibration slope},\\
0, & \text{for calibration intercept and CITL}.
\end{cases}
\label{eq:calibration-diagonal}
\end{equation}
This identity does not require correct logistic specification. Because the value in Equation~\eqref{eq:calibration-diagonal} is constant for each calibration measure, $\Omega_{H,n}^{\mathrm{cal}}(P)=\Psi_{\mathrm{cal}}(P,P)-C_{\mathrm{proc},n}^{\mathrm{cal}}(P).$ Together with the definition of $C_{H,n}$ in Equation~\eqref{eq:point-estimators}, this gives $C_{H,n}^{\mathrm{cal}}(P)=C_{\mathrm{proc},n}^{\mathrm{cal}}(P).$

Because $\Psi_{\mathrm{cal}}(P,P)$ remains constant as $P$ changes, differentiating Equation~\eqref{eq:calibration-diagonal} gives
\begin{equation}
\phi_{\mathrm{tr}}^{\mathrm{cal}}
+
\phi_{\mathrm{ev}}^{\mathrm{cal}}
=
0
\qquad P_0\text{-a.s.}
\label{eq:calibration-if-cancellation}
\end{equation}
In Equation~\eqref{eq:generic-estimand-roots}, the leading term of the estimation error for $\theta_{\mathrm{proc},n}^{\mathrm{cal}}$ contains this sum, whereas that for $\theta_{\mathrm{FM},n}^{\mathrm{cal}}$ contains only $\phi_{\mathrm{ev}}^{\mathrm{cal}}$. Substitution therefore gives
\begin{equation}
\begin{aligned}
\sqrt n\{C_{H,n}^{\mathrm{cal}}(P_n)-\theta_{\mathrm{FM},n}^{\mathrm{cal}}\}
&=\G_n\phi_{\mathrm{ev}}^{\mathrm{cal}}+o_p(1),\\
\sqrt n\{C_{H,n}^{\mathrm{cal}}(P_n)-\theta_{\mathrm{proc},n}^{\mathrm{cal}}\}
&=o_p(1).
\end{aligned}
\label{eq:calibration-root-separation}
\end{equation}
Thus, the root for $\theta_{\mathrm{proc},n}^{\mathrm{cal}}$ is degenerate on the $\sqrt n$ scale, whereas the root for $\theta_{\mathrm{FM},n}^{\mathrm{cal}}$ remains nondegenerate when $\Var_{P_0}(\phi_{\mathrm{ev}}^{\mathrm{cal}})>0$. The corrected point estimate is centered at $c_0$ to first order, because $C_{H,n}^{\mathrm{cal}}(P_n)-c_0=o_p(n^{-1/2})$, where $c_0=1$ for calibration slope and $c_0=0$ for calibration intercept and CITL. Even so, the interval targets the population calibration of the realized fitted model, $\theta_{\mathrm{FM},n}^{\mathrm{cal}}$. Since $\sqrt n(\theta_{\mathrm{FM},n}^{\mathrm{cal}}-c_0)
=-\G_n\phi_{\mathrm{ev}}^{\mathrm{cal}}+o_p(1)$, its width reflects the sampling variation of this random target around $c_0$, and its coverage refers to repeated development samples.

For the calibration measures, Equations~\eqref{eq:method-root-map} and~\eqref{eq:calibration-if-cancellation} give
\[
\sqrt n\{D_b^{\mathcal M}-C_{H,n}^{\mathrm{cal}}(P_n)\}
=
\begin{cases}
-\G_n^\star\phi_{\mathrm{ev}}^{\mathrm{cal}}+o_{p^\star}(1),
  & \mathcal M\in\{\mathrm{NEB},\mathrm{OEB}\},\\
o_{p^\star}(1),
  & \mathcal M\in\{\mathrm{TSB},\mathrm{LSB},\mathrm{bTSB},\mathrm{bLSB}\}.
\end{cases}
\]
Although TSB recalculates the correction in each outer bootstrap sample, it fails to capture the first-order variation required for inference on $\theta_{\mathrm{FM},n}^{\mathrm{cal}}$. Nonzero interval widths for TSB and bTSB can arise from smaller-order variation in the correction or from Monte Carlo error due to finite inner resampling. Apparent, LSB, and bLSB yield intervals of zero width.

\section{Application to the GUSTO-I West region data}\label{sec:application}

\subsection{Analysis}\label{sec:application-methods}

To illustrate differences among the CIs in real data, we applied Apparent, NEB, OEB, TSB, and LSB to the GUSTO-I West region data set. GUSTO-I was an international randomized trial comparing four thrombolytic strategies for acute myocardial infarction \citep{gusto1993}.  Following \citet{noma2021}, we used death within 30 days as the outcome. The analysis included 2188 participants, of whom 135 died within 30 days (event fraction, 0.0617). We fitted logistic regression models by maximum likelihood without penalization using the prespecified sets of 8 and 17 predictors, corresponding to 16.9 and 7.94 events per variable (EPV), respectively. Definitions and coding of the predictors are given in Supporting Information Section~S5.1 \citep{noma2021}. We evaluated AUC, Brier score, log loss, calibration slope, calibration intercept, and CITL. We focused on the calibration measures, for which NEB and OEB retain nondegenerate first-order variation for $\theta_{\mathrm{FM},n}^{\mathrm{cal}}$, whereas TSB and LSB are degenerate on the $\sqrt n$ scale, as shown in Equations~\eqref{eq:method-root-map} and~\eqref{eq:calibration-root-separation}. We used 1000 outer bootstrap samples for all CIs and 100 inner samples per outer sample for the nested methods. For comparison, we also calculated a DeLong CI for the apparent AUC \citep{delong1988}.

\subsection{Results}\label{sec:application-results}

Table~\ref{tab:gusto-real-data} summarizes the application results. The apparent AUC was 0.819 for the 8-predictor model and 0.832 for the 17-predictor model. After Harrell's bias correction, the estimates were 0.809 and 0.811, respectively. NEB and OEB yielded similar CIs for AUC, Brier score, and log loss. For AUC, the NEB and OEB intervals had widths of 0.077 and 0.069 for the 8-predictor model and 0.072 and 0.068 for the 17-predictor model, respectively.

The calibration measures showed a different pattern. NEB and OEB yielded substantially wider intervals than TSB, whereas Apparent and LSB yielded intervals of zero width. For the 17-predictor model, the TSB interval for calibration slope was narrow and excluded the ideal value of 1 (0.877, 0.924), whereas the NEB interval was much wider and included 1 (0.755, 1.053). For calibration intercept, the TSB interval excluded the ideal value of 0 (-0.274, -0.141), whereas the NEB interval included 0 (-0.593, 0.215). For CITL, both the TSB interval (-0.017, 0.024) and the NEB interval (-0.205, 0.202) included 0, but the NEB interval was substantially wider. The corrected point estimates were the same for these methods, but the reported uncertainty about the calibration of the fitted model differed greatly. This illustrates the importance of matching the bootstrap interval to the estimand being reported.

\section{Simulation study}\label{sec:simulation}

\subsection{Simulation settings}\label{sec:simulation-settings}

The simulation study comprised 21 GUSTO-I-based scenarios (Simulation 1) and 49 synthetic scenarios (Simulation 2). The design and reporting followed the framework of aims, data-generating mechanisms, estimands, methods, and performance measures \citep{morris2019}. For both simulations, the development sample size was set to $n=\lceil p\mathrm{EPV}/\text{event fraction}\rceil$, where $p$ is the prespecified number of predictor variables.

\noindent\textbf{Aims:} The aims were to evaluate the coverage of 95\% CIs based on NEB and OEB for $\theta_{\mathrm{FM},n}$, to compare them with methods that retain the first-order variation for $\theta_{\mathrm{proc},n}$, and to assess the approximation of NEB by OEB across performance measures and EPV.

\noindent\textbf{Data-generating mechanisms for Simulation 1 (GUSTO-I-based settings).} We considered three settings based on the GUSTO-I West region data \citep{gusto1993}. These settings were constructed to approximate the predictor distributions and dependence structure observed in the data while varying the predictor set, event fraction, and EPV. The data-generating mechanism followed \citet{noma2021}. We used the same complete-case data set as in the application (Section~\ref{sec:application}) to estimate the predictor distributions and the coefficients of the logistic model. Age, height, and weight were generated jointly from a multivariate normal distribution. Smoking was generated from fitted multinomial probabilities. The remaining binary predictors were generated from a multivariate Bernoulli distribution fitted by iterative proportional fitting. The predictor coefficients were fixed at the maximum likelihood estimates from logistic regression fitted without penalization. For each target event fraction $\pi$, only the intercept $\alpha_\pi$ was recalibrated to satisfy $\E_{\widehat F_X}[\expit\{\alpha_\pi+\widehat\beta^{\mathsf T}X\}]=\pi$, where $\widehat F_X$ is the fitted distribution of the predictors. Outcomes were then generated from the resulting Bernoulli model. The three settings used (i) 8 predictors and event fraction 0.125, (ii) 8 predictors and event fraction 0.0625, and (iii) 17 predictors and event fraction 0.125. Each setting used EPV 1, 3, 5, 7, 10, 20, and 40. Definitions and coding of the predictors are given in Supporting Information Section~S5.1.

\noindent\textbf{Data-generating mechanisms for Simulation 2 (synthetic settings).} We considered five correctly specified synthetic settings, based in part on design factors used by \citet{vanSmeden2019}. The baseline setting used eight independent Bernoulli$(0.5)$ predictors and population AUC 0.75. The other settings used eight independent standard normal predictors, population AUC 0.65 or 0.85, or eight Bernoulli$(0.5)$ predictors with pairwise Pearson correlation 0.5. In each correctly specified scenario, the relative regression coefficients were fixed and multiplied by a common scale. This scale was selected numerically so that a Monte Carlo approximation to the population AUC matched the prespecified value, and the intercept was selected to give an event fraction of 0.125. Details are given in Supporting Information Section~S5.2.

We also examined model misspecification using eight independent standard normal predictors. We added either a quadratic term, $\gamma(X_1^2-1)/\sqrt{2}$, or an interaction term, $\gamma X_1X_2$, to the log-odds of the outcome, with $\gamma=1$. We retained the linear coefficients from the correctly specified setting with population AUC 0.75 and adjusted the intercept to maintain an event fraction of 0.125. The added term was omitted from the fitted logistic regression model.

All settings used eight predictors, a target event fraction of 0.125, and EPV 1, 3, 5, 7, 10, 20, and 40. The corresponding development sample sizes were 64, 192, 320, 448, 640, 1280, and 2560. We restricted the distributional diagnostics (Figure~\ref{fig:distributional-diagnostic}) and the comparison of NEB and OEB (Table~\ref{tab:neb-oeb-approximation}) to the baseline synthetic setting.

\noindent\textbf{Estimand:} The primary estimand was $\theta_{\mathrm{FM},n}=\Psi(P_n,P_0)$. For each simulation replicate, $\theta_{\mathrm{FM},n}$ was evaluated using an independent population sample of 200,000 observations. We considered six performance measures: AUC, Brier score, log loss, calibration slope, calibration intercept, and CITL.

\noindent\textbf{Methods:} In all scenarios, the development model included all prespecified predictors and was fitted by logistic regression using maximum likelihood without penalization. Apparent, NEB, OEB, TSB, and LSB were evaluated in all simulations. NEB and OEB were used for inference on $\theta_{\mathrm{FM},n}$. Apparent was included as an uncorrected baseline. TSB and LSB were included as the methods that retain the first-order variation for $\theta_{\mathrm{proc},n}$. In Simulation 1, bTSB and bLSB were compared with TSB and LSB, respectively, to assess the effect on coverage of using percentile versus basic bootstrap intervals while retaining the same first-order variation for $\theta_{\mathrm{proc},n}$. 

We used 1000 outer bootstrap samples for all methods. Nested methods used $B_{\mathrm{inner}}=100$ inner bootstrap samples. Before the final simulation, a prespecified pilot evaluated $B_{\mathrm{inner}}=50,100,200,400,600,800,$ and $1000$ and selected $B_{\mathrm{inner}}=100$. Coverage and interval width were not used in this choice; selection was based only on the Monte Carlo precision of the inner bootstrap. Details are given in Supporting Information Section~S4.1.

\noindent\textbf{Performance measures:} Each scenario used 1000 simulation replicates. The primary performance measure was coverage of the 95\% CI for $\theta_{\mathrm{FM},n}$. For each method and performance measure, we considered a simulation replicate successful if a CI was obtained using estimates from converged fits in at least 90\% of outer bootstrap samples and, when used, at least 90\% of inner samples within each retained outer sample. We reported coverage among successful replicates with an available target value, together with the success rate. When all 1000 replicates contribute to the coverage estimate, the Monte Carlo standard error is approximately 0.007 at 95\% coverage. We also reported the median and the 90th and 95th percentiles of CI width among successful replicates. Coverage and CI width summarize interval performance but do not characterize agreement between each bootstrap distribution and the sampling distribution of the estimation error for $\theta_{\mathrm{FM},n}$. We therefore assessed this agreement using the probability integral transform (PIT) and the first Wasserstein distance $W_1$. Because each bootstrap distribution was represented by a finite number of replicates, we used the mid-rank form of the PIT, which gives half weight to ties and remains strictly between 0 and 1 \citep{sailynoja2022}. If the bootstrap distribution approximates the relevant sampling distribution, the PIT is approximately uniform across simulation replicates \citep{gneitingBalabdaouiRaftery2007}. We used the transformation $Z_{\mathrm{PIT}}=\Phi^{-1}(\mathrm{PIT})$, where $\Phi$ is the standard normal cumulative distribution function. A mean close to 0 and a standard deviation (SD) close to 1 indicate good agreement.

We also used $W_1$ to compare each bootstrap distribution with the corresponding sampling distribution for $\theta_{\mathrm{FM},n}$ \citep{peyreCuturi2019}. For each simulation replicate, the sampling distribution was estimated from the remaining replicates. We divided $W_1$ by the SD of the reference root for $\theta_{\mathrm{FM},n}$ and used the same denominator for all methods within each scenario and performance measure. Full definitions are given in Supporting Information Section~S4.2.  For the approximation of NEB by OEB, we write $\Delta\Omega_H$ for the collection of outer-bootstrap values $\{\Delta\Omega_{H,b}\}_{b=1}^{B_{\mathrm{outer}}}$ defined in Section~\ref{sec:neb-oeb}. We additionally examined the ratio of the median interval width for OEB to that for NEB, $\sqrt n\,\operatorname{RMSE}_\star(\Delta\Omega_H)/\operatorname{SD}_\star(R_{\mathrm{FM},n}^\star)$, and $\sqrt n\,\operatorname{RMSE}_\star(\Delta\Omega_H)$.

\subsection{Simulation results}\label{sec:simulation-results}

At EPV 1, the minimum success rates for constructing CIs were 82.7\% in Simulation 1 and 80.5\% in Simulation 2. At EPV 3 or higher, all methods achieved 100\% success across the reported settings and performance measures (Supporting Information Figure S4).

\subsubsection{Coverage in Simulation 1 (GUSTO-I-based settings)}

Coverage in Simulation 1 is shown in Figure~\ref{fig:gusto-coverage}. For AUC, NEB was generally closer to the nominal level at low EPV, and OEB approached NEB as EPV increased. The difference between NEB and OEB was largest at low EPV, particularly in the setting with 17 predictors. Apparent, TSB, LSB, bTSB, and bLSB generally showed greater undercoverage for AUC at low EPV. For Brier score and log loss, NEB and OEB yielded similar coverage, and differences between them were small across most EPV values. Apparent, TSB, LSB, bTSB, and bLSB showed greater undercoverage at low EPV, with the differences decreasing as EPV increased.

The calibration measures showed clear differences among the methods. For calibration slope and intercept, coverage varied substantially at low EPV and was generally lower for OEB than for NEB. At EPV 10 or higher, coverage for both NEB and OEB was generally close to the nominal level. For CITL, NEB and OEB showed overcoverage at EPV 1. At EPV 3 or higher, their coverage was generally close to the nominal level. Across all three calibration measures, TSB and bTSB showed substantial undercoverage and coverage for Apparent, LSB, and bLSB was near zero at EPV 3 or higher.

\subsubsection{Coverage in Simulation 2 (synthetic settings)}

In Simulation 2, the relative patterns among the methods were similar across changes in predictor type, population AUC, and correlation and in the two misspecified settings (Figure~\ref{fig:synthetic-coverage}). For AUC, OEB showed greater undercoverage than NEB at low EPV and approached NEB more gradually than for Brier score or log loss. For Brier score and log loss, NEB and OEB yielded similar coverage across most EPV values. For Brier score, the first-order training contribution $\phi_{\mathrm{tr}}$ need not vanish under model misspecification. At EPV 10, 20, and 40, however, the observed changes in coverage were small: under either form of misspecification, coverage for each of NEB, OEB, TSB, and LSB differed by at most 2.0 percentage points from that in the correctly specified setting with continuous predictors. For AUC, $\phi_{\mathrm{tr}}$ also need not vanish under model misspecification. Nevertheless, differences in coverage among these four methods were small at higher EPV in the two misspecified settings.

The calibration measures again showed clear differences among the methods. For calibration slope and intercept, coverage for NEB and OEB generally approached the nominal level as EPV increased, although undercoverage persisted in some settings, particularly for OEB when the population AUC was 0.65. At EPV 40, coverage ranged from 93.3\% to 95.5\% for NEB and from 89.9\% to 95.3\% for OEB across these two measures and all settings. For CITL, NEB and OEB generally yielded coverage close to the nominal level at EPV 3 or higher. Across all three calibration measures, TSB coverage generally decreased as EPV increased, reaching approximately 15--20\% at EPV 40. Coverage for Apparent and LSB was near zero at EPV 3 or higher.

\subsubsection{Distributional diagnostics in Simulation 2}
\label{sec:results-distributional-diagnostic}

Figure~\ref{fig:distributional-diagnostic} shows the distributional diagnostics in the baseline setting of Simulation 2. For AUC, Brier score, and log loss, NEB generally yielded a mean $Z_{\mathrm{PIT}}$ closer to 0, a SD closer to 1, and a smaller standardized $W_1$. OEB approached NEB as EPV increased. The differences became small by EPV 5 for Brier score and log loss but decreased more gradually for AUC. 

For the calibration measures, the methods showed larger differences in CI width and distributional agreement. Across all three measures, NEB yielded an SD of $Z_{\mathrm{PIT}}$ close to 1 and small standardized $W_1$ at EPV 10--40. For calibration slope and intercept, OEB approached NEB in these diagnostics as EPV increased, whereas the two methods yielded similar values for CITL at EPV 3 or higher. Median CI widths for calibration slope and intercept were smaller for OEB than for NEB at EPV 10 or lower. Across all three measures, median CI widths for TSB decreased toward zero as EPV increased, whereas those for LSB and Apparent were near zero at EPV 3 or higher. For TSB, LSB, and Apparent, the SD of $Z_{\mathrm{PIT}}$ remained substantially above 1 and standardized $W_1$ remained large even at higher EPV.

\subsubsection{Approximation of NEB by OEB in Simulation 2}
\label{sec:results-fixed-correction}

Table~\ref{tab:neb-oeb-approximation} shows the approximation of NEB by OEB in the baseline setting of Simulation 2. For AUC, the OEB/NEB width ratio increased from 0.774 at EPV 1 to 0.965 at EPV 40. For Brier score and log loss, the width ratios were already close to 1 at EPV 5 (0.968 and 1.02, respectively) and remained close to 1 at higher EPV. For calibration slope and intercept, the width ratios approached 1 more gradually, whereas the ratio for CITL remained close to 1 across all EPV values. At EPV 40, $\sqrt n\,\operatorname{RMSE}_\star(\Delta\Omega_H)/\operatorname{SD}_\star(R_{\mathrm{FM},n}^\star)$ was approximately 0.1 across all six measures.

\section{Discussion}\label{sec:discussion}

We developed an estimand-specific framework for CIs after bootstrap optimism correction. The performance of the final fitted model, $\theta_{\mathrm{FM},n}$, is often the estimand of interest when evaluating prediction models. For inference on $\theta_{\mathrm{FM},n}$, we developed NEB as a new reference method and derived OEB as a computationally simpler approximation. Our theory showed that NEB represents the first-order variation required for $\theta_{\mathrm{FM},n}$. In contrast, TSB, LSB, bTSB, and bLSB reproduce the first-order variation associated with $\theta_{\mathrm{proc},n}$, rather than the estimation error for $\theta_{\mathrm{FM},n}$. Under the regularity conditions considered here, this variation agrees with that required for first-order inference on $\theta_{\mathrm{proc},n}$. These methods therefore do not in general provide the distribution required for inference on $\theta_{\mathrm{FM},n}$. This mismatch between the estimands had a clear consequence for the calibration measures, which are important for evaluating clinical prediction models \citep{vanCalster2019,collinsEvaluation2024,riley2021,sadatsafavi2024}. For the calibration measures, the first-order root for $\theta_{\mathrm{proc},n}$ is degenerate on the $\sqrt n$ scale, whereas that for $\theta_{\mathrm{FM},n}$ has nonzero first-order variation. The simulation results were consistent with this distinction. The resulting CIs depended not only on the optimism-corrected point estimate but also on the first-order variation represented by the bootstrap root. Thus, we recommend choosing the bootstrap distribution according to the estimand and performance measure being reported.

For $\theta_{\mathrm{FM},n}$, NEB provides a reference method but requires an inner bootstrap within each outer sample. OEB removes this inner loop and is computationally simpler. One-level constructions algebraically equivalent to OEB have been used previously for AUC and calibration measures \citep{baart2019,meisner2020,martin2021}. Our derivation identifies OEB as an approximation to NEB and shows that their exact paired difference is $\Delta\Omega_H$, the change in Harrell's bias correction. Agreement between NEB and OEB improved rapidly for Brier score and log loss but more gradually for AUC. OEB is computationally attractive, but its finite-sample behavior should not be assumed to match that of NEB, particularly for the calibration measures at low EPV (Table 2). For the calibration measures, $\Delta\Omega_H$ also characterizes the finite-sample variation remaining in TSB: the deviation of the TSB replicate from the corrected estimate is $-\Delta\Omega_H$. Thus, narrow but nonzero TSB intervals may reflect variation in $\Delta\Omega_H$ across outer bootstrap samples rather than the first-order variation required for inference on $\theta_{\mathrm{FM},n}$. In addition, Monte Carlo error from finite inner resampling can contribute to the width of TSB intervals. Consistent with this interpretation, smaller $B_{\mathrm{inner}}$ yielded wider TSB intervals and higher coverage for the calibration measures in the sensitivity analysis (Supporting Information Section~S4.1 and Figures~S1 and~S2).

The effect of using a bootstrap distribution for a different estimand depends on the performance measure. Under correct logistic specification, $\phi_{\mathrm{tr}}=0$ for log loss and Brier score when the model is fitted by maximum likelihood without penalization. The population linear predictor $\eta_{P_0}$ ranks observations by their conditional event probabilities $P_0(Y=1\mid X)$ and therefore maximizes population AUC \citep{mcintoshPepe2002}. If population AUC is differentiable with respect to the regression coefficients at their true values, its gradient is zero at that point and $\phi_{\mathrm{tr}}=0$ for AUC as well (Supporting Information Section~S3). Under these conditions, the estimation errors for $\theta_{\mathrm{FM},n}$ and $\theta_{\mathrm{proc},n}$ agree to first order for log loss, Brier score, and AUC. In our simulations, differences in AUC coverage among NEB, OEB, TSB, and LSB were small at higher EPV, consistent with the favorable coverage that Noma et al.\ reported for the C-statistic \citep{noma2021}. Under model misspecification, $\phi_{\mathrm{tr}}$ remains zero for log loss because the limiting coefficients still minimize expected log loss. For Brier score and AUC, $\phi_{\mathrm{tr}}$ can be nonzero. At EPV 10 or higher, however, each of the four methods yielded similar Brier score coverage in the correctly specified setting with continuous predictors and in the two misspecified settings. For the three calibration measures, the identity $\phi_{\mathrm{tr}}^{\mathrm{cal}}+\phi_{\mathrm{ev}}^{\mathrm{cal}}=0$ holds under both correct specification and misspecification (Equation~\eqref{eq:calibration-if-cancellation}). Thus, on the $\sqrt n$ scale, the estimation error for $\theta_{\mathrm{proc},n}^{\mathrm{cal}}$ converges to zero, whereas that for $\theta_{\mathrm{FM},n}^{\mathrm{cal}}$ has a nondegenerate limit. To our knowledge, these contrasting limits have not previously been derived in terms of the training and evaluation influence functions. This derivation explains why Martin et al.\ obtained nondegenerate CIs for calibration slope and CITL using a procedure algebraically equivalent to OEB \citep{martin2021}.

Several limitations should be considered. Our theory assumes a fixed number of predictors. Under the stated conditions, the difference between the corrected and apparent estimators is $o_p(n^{-1/2})$. At low EPV, higher-order terms may contribute to differences in coverage among methods, and our first-order approximation cannot quantify their contribution. In addition, the simulations included a limited range of predictor distributions, event fractions, population AUCs, correlation structures, and model misspecification. Coverage was evaluated among replicates with successful CI construction and an available target value. At EPV 1, this restriction may affect comparisons among methods, so coverage should be interpreted together with the CI construction success rates (Supporting Information Figure~S4). For calibration, we evaluated the slope, intercept, and CITL but did not assess the full calibration curve \citep{vanCalster2016,sadatsafavi2024}. Furthermore, we focused on Harrell's bias correction and did not evaluate the .632 or .632+ estimators \citep{efron1983,efronTibshirani1997}. Moreover, some settings with binary predictors do not meet the sufficient conditions used to derive our theoretical results for AUC. In these settings, different combinations of the predictors’ values yield the same score at the population level. We therefore rely on simulation to assess the performance of CIs in these settings. Finally, because nested resampling requires substantial computation, we restricted this study to unpenalized logistic regression with a fixed number of parameters. As penalization, tuning, and variable selection may change the first-order structure, future work should extend the theory to procedures such as LASSO and develop suitable bootstrap CIs for these settings.

\section{Conclusion}\label{sec:conclusion}

We developed an estimand-specific framework for bootstrap CIs after optimism correction. For inference on $\theta_{\mathrm{FM},n}$, NEB provides a nested reference method, whereas OEB provides a computationally simpler approximation when the change in Harrell's bias correction is negligible on the $\sqrt n$ scale. The distinction between the estimands was particularly important for the calibration measures, for which TSB and LSB have degenerate first-order variation while NEB and OEB retain the variation required for inference on $\theta_{\mathrm{FM},n}$. Bootstrap CIs after optimism correction should therefore be constructed with regard to both the estimand and the performance measure.

\appendixtitleon
\begin{appendices}
\numberwithin{equation}{section}

\section{Cancellation of first-order variation for calibration measures}
\label{app:calibration}

In this appendix, we define the calibration measures as functions of $P$ and $Q$ and derive the cancellation of their first-order variation. As in the main text, $P$ denotes the distribution used to fit the logistic model and $Q$ the distribution used to evaluate its calibration. We assume throughout this appendix that the required logistic and calibration estimating equations have finite, locally unique solutions. We use the first-order expansions and conditions for the corrected point estimators stated in Supporting Information Section~S1.

Let $\widetilde X=(1,X^{\mathsf T})^{\mathsf T}$ and define $p_\beta(X)=\expit(\widetilde X^{\mathsf T}\beta)$. Let $\beta(P)$ denote the logistic regression functional defined by
\[
P\!\left[
\widetilde X\{Y-p_\beta(X)\}
\right]=0.
\]
Write $\eta_P(X)=\widetilde X^{\mathsf T}\beta(P)$. The calibration intercept and slope, denoted by $\alpha_{\mathrm{cal}}(P,Q)$ and $\gamma_{\mathrm{cal}}(P,Q)$, are defined jointly as the solution $(\alpha,\gamma)$ of
\begin{equation}
Q\!\left[
\begin{pmatrix}
1\\
\eta_P(X)
\end{pmatrix}
\left\{
Y-\expit\bigl(\alpha+\gamma\eta_P(X)\bigr)
\right\}
\right]=0.
\label{eqA:recalibration-score}
\end{equation}
For CITL, the slope is fixed at 1. The intercept $\alpha_{\mathrm{CITL}}(P,Q)$ is the solution in $\alpha$ of
\[
Q\!\left[
Y-\expit\bigl(\alpha+\eta_P(X)\bigr)
\right]=0.
\]
We write $\Psi_{\mathrm{cal}}(P,Q)$ for any one of these three measures. When $Q=P$, $\alpha=0$ solves the equation for CITL and $(\alpha,\gamma)=(0,1)$ solves Equation~\eqref{eqA:recalibration-score}. For the slope equation,
\[
P\!\left[
\eta_P(X)\{Y-p_{\beta(P)}(X)\}
\right]
=
\beta(P)^{\mathsf T}
P\!\left[
\widetilde X\{Y-p_{\beta(P)}(X)\}
\right]
=0.
\]
Hence, under local uniqueness,
\begin{equation}
\Psi_{\mathrm{cal}}(P,P)=c_0,
\label{eqA:calibration-diagonal}
\end{equation}
where $c_0=1$ for calibration slope and $c_0=0$ for calibration intercept and CITL. This identity does not require correct logistic specification. Because Equation~\eqref{eqA:calibration-diagonal} holds for every such $P$,
\begin{equation}
\Omega_{H,n}^{\mathrm{cal}}(P)
=
c_0-C_{\mathrm{proc},n}^{\mathrm{cal}}(P),
\qquad
C_{H,n}^{\mathrm{cal}}(P)
=
C_{\mathrm{proc},n}^{\mathrm{cal}}(P).
\label{eqA:calibration-center-equality}
\end{equation}
Thus, the estimators $C_{H,n}^{\mathrm{cal}}$ and $C_{\mathrm{proc},n}^{\mathrm{cal}}$ coincide exactly.

For local identification, let $\eta_0=\eta_{P_0}$, $p_0=\expit(\eta_0)$, and $v_0=p_0(1-p_0)$. The negative Jacobian of the estimating function in Equation~\eqref{eqA:recalibration-score} at $(0,1)$ is
\[
M_0=
\begin{pmatrix}
P_0v_0 & P_0(v_0\eta_0)\\
P_0(v_0\eta_0) & P_0(v_0\eta_0^2)
\end{pmatrix}.
\]
It is nonsingular if
\[
P_0v_0>0,
\qquad
P_0\!\left[
v_0\{\eta_0-\mu_v\}^2
\right]>0,
\qquad
\mu_v=\frac{P_0(v_0\eta_0)}{P_0v_0}.
\]
CITL requires only $P_0v_0>0$. Together with local pathwise differentiability of $\beta(P)$ and suitable moment conditions, these identification conditions give local pathwise differentiability of the calibration parameters by the implicit function theorem.

Let $\mathcal T_{P_0}$ denote the tangent space at $P_0$, defined as the closed linear span in $L_2(P_0)$ of the scores of regular one-dimensional parametric submodels through $P_0$. For any such submodel score $g$, differentiation of Equation~\eqref{eqA:calibration-diagonal} at $t=0$ gives
\[
0
=
\left.
\frac{d}{dt}
\Psi_{\mathrm{cal}}(P_t,P_t)
\right|_{t=0}
=
P_0\!\left[
\left\{
\phi_{\mathrm{tr}}^{\mathrm{cal}}
+
\phi_{\mathrm{ev}}^{\mathrm{cal}}
\right\}g
\right].
\]
By linearity and continuity, this relation extends to every $g\in\mathcal T_{P_0}$. For the unrestricted nonparametric model, $\mathcal T_{P_0}=L_2^0(P_0)$. Since $\phi_{\mathrm{tr}}^{\mathrm{cal}}+\phi_{\mathrm{ev}}^{\mathrm{cal}}\in L_2^0(P_0)$, taking $g=\phi_{\mathrm{tr}}^{\mathrm{cal}}+\phi_{\mathrm{ev}}^{\mathrm{cal}}$ gives $P_0\{(\phi_{\mathrm{tr}}^{\mathrm{cal}}+\phi_{\mathrm{ev}}^{\mathrm{cal}})^2\}=0$, and hence
\begin{equation}
\phi_{\mathrm{tr}}^{\mathrm{cal}}
+
\phi_{\mathrm{ev}}^{\mathrm{cal}}
=
0
\qquad
P_0\text{-a.s.}
\label{eqA:calibration-cancellation}
\end{equation}
Under the conditions in Supporting Information Section~S1, Equation~\eqref{eqA:calibration-cancellation} gives the following first-order relations:
\begin{equation}
\begin{aligned}
\sqrt n
\{C_{H,n}^{\mathrm{cal}}(P_n)
-\theta_{\mathrm{FM},n}^{\mathrm{cal}}\}
&=
\G_n\phi_{\mathrm{ev}}^{\mathrm{cal}}+o_p(1),\\
\sqrt n
\{C_{H,n}^{\mathrm{cal}}(P_n)
-\theta_{\mathrm{proc},n}^{\mathrm{cal}}\}
&=
o_p(1).
\end{aligned}
\label{eqA:calibration-root-separation}
\end{equation}
By Equation~\eqref{eqA:calibration-center-equality}, the second relation also holds for $C_{\mathrm{proc},n}^{\mathrm{cal}}(P_n)$. Thus, the root for $\theta_{\mathrm{proc},n}^{\mathrm{cal}}$ is degenerate on the $\sqrt n$ scale, whereas the root for $\theta_{\mathrm{FM},n}^{\mathrm{cal}}$ is nondegenerate when $\Var_{P_0}(\phi_{\mathrm{ev}}^{\mathrm{cal}})>0$. Applying Equation~\eqref{eq:method-root-map} under the corresponding conditions in Supporting Information Section~S2.2 gives
\begin{equation}
\sqrt n
\{D_{b,\mathrm{cal}}^{\mathcal M}
-C_{H,n}^{\mathrm{cal}}(P_n)\}
=
\begin{cases}
-\G_n^\star\phi_{\mathrm{ev}}^{\mathrm{cal}}
+o_{p^\star}(1),
&
\mathcal M\in\{\mathrm{NEB},\mathrm{OEB}\},\\[3pt]
o_{p^\star}(1),
&
\mathcal M\in\{\mathrm{TSB},\mathrm{LSB},
\mathrm{bTSB},\mathrm{bLSB}\}.
\end{cases}
\label{eqA:calibration-method-map}
\end{equation}
The common centering value in Equation~\eqref{eqA:calibration-method-map} follows from Equation~\eqref{eqA:calibration-center-equality}. In addition,
\[
D_{b,\mathrm{cal}}^{\mathrm{App}}=c_0,
\qquad
D_{b,\mathrm{cal}}^{\mathrm{LSB}}
=
D_{b,\mathrm{cal}}^{\mathrm{bLSB}}
=
C_{H,n}^{\mathrm{cal}}(P_n).
\]
Thus, the Apparent, LSB, and bLSB distributions are exactly degenerate. For TSB and bTSB, Equation~\eqref{eqA:calibration-center-equality} gives
\begin{equation}
D_{b,\mathrm{cal}}^{\mathrm{bTSB}}
=
2C_{H,n}^{\mathrm{cal}}(P_n)
-
D_{b,\mathrm{cal}}^{\mathrm{TSB}}.
\label{eqA:calibration-tsb-btsb}
\end{equation}
Thus, when the conditional inner-bootstrap expectation is used, their bootstrap distributions are reflections about $C_{H,n}^{\mathrm{cal}}(P_n)$. With the same outer bootstrap samples and quantile rule, their equal-tailed intervals have the same width, although their lower and upper confidence limits need not coincide. For TSB defined using the conditional inner-bootstrap expectation,
\begin{equation}
D_{b,\mathrm{cal}}^{\mathrm{TSB}}
-
C_{H,n}^{\mathrm{cal}}(P_n)
=
-\Delta\Omega_{H,b}^{\mathrm{cal}}.
\label{eqA:tsb-calibration}
\end{equation}
Hence, $\sqrt n\,\Delta\Omega_{H,b}^{\mathrm{cal}}=o_{p^\star}(1)$ implies that TSB is degenerate on the $\sqrt n$ scale. If both confidence limits are $C_{H,n}^{\mathrm{cal}}(P_n)+o_p(n^{-1/2})$ and the root for $\theta_{\mathrm{FM},n}^{\mathrm{cal}}$ has a continuous nondegenerate limit, the coverage probability for $\theta_{\mathrm{FM},n}^{\mathrm{cal}}$ converges to zero. Nonzero finite-sample widths of TSB and bTSB can arise from smaller-order variation, numerical irregularity, or Monte Carlo error from finite inner resampling and do not represent the first-order variation for $\theta_{\mathrm{FM},n}^{\mathrm{cal}}$.

\end{appendices}

\clearpage
\section*{Declarations}

\subsection*{Ethics statement}

Ethics approval was not required because this study involved simulation studies and secondary analysis of an existing publicly available data set.

\subsection*{Funding}

This research was supported by AMED under Grant Number JP266k0237002.

\subsection*{Conflict of interest}

The authors declare no conflicts of interest.

\subsection*{Author contributions}

S.T.: Conceptualization, Methodology, Software, Formal analysis, Investigation, Visualization, Writing -- original draft, Writing -- review \& editing. K.N.: Writing -- review \& editing. R.E.: Writing -- review \& editing. S.M.: Writing -- review \& editing.

\subsection*{Data availability statement}

The GUSTO-I West region data set is available through the Clinical Prediction Models website (www.clinicalpredictionmodels.org). R code for all methods, simulations, and the application is available from the corresponding author upon reasonable request.

\subsection*{Acknowledgments}

None.

\subsubsection*{Declaration of generative AI and AI-assisted technologies in the manuscript preparation process}

During the preparation of this work, the authors used ChatGPT to assist with language refinement, editing suggestions, and improvements in clarity and readability. After using this tool, the authors reviewed and edited the content as needed and take full responsibility for the content of the published article.

\subsection*{Supporting information}

The Supporting Information is included after the main article in this preprint.

\subsection*{ORCID}

Seiya Takano: \url{https://orcid.org/0000-0001-6990-9923}

\clearpage
\renewcommand{\refname}{References}
\bibliographystyle{oup-abbrvnat}
\bibliography{references}

\clearpage
\begingroup
\footnotesize
\setlength{\tabcolsep}{3pt}
\renewcommand{\arraystretch}{1.04}
\begin{longtable}{@{}>{\raggedright\arraybackslash}p{0.15\textwidth} >{\raggedright\arraybackslash}p{0.13\textwidth} >{\raggedleft\arraybackslash}p{0.30\textwidth} >{\raggedleft\arraybackslash}p{0.30\textwidth}@{}}
\caption{Real-data illustration using complete cases from the GUSTO-I West region data set.}
\label{tab:gusto-real-data}\\
\toprule
Performance measure & Method & 8-predictor model & 17-predictor model \\
\midrule
\endfirsthead
\multicolumn{4}{l}{\textit{Table \thetable\ continued}}\\
\toprule
Performance measure & Method & 8-predictor model & 17-predictor model \\
\midrule
\endhead
\bottomrule
\endfoot
\endlastfoot

\multirow[t]{6}{0.15\textwidth}{AUC}
& Apparent & 0.819 (0.789, 0.858) & 0.832 (0.805, 0.876) \\
& NEB      & 0.809 (0.774, 0.851) & 0.811 (0.776, 0.848) \\
& OEB      & 0.809 (0.776, 0.845) & 0.811 (0.777, 0.845) \\
& TSB      & 0.809 (0.774, 0.850) & 0.811 (0.783, 0.856) \\
& LSB      & 0.809 (0.780, 0.849) & 0.811 (0.785, 0.856) \\
& DeLong   & 0.819 (0.783, 0.854) & 0.832 (0.796, 0.867) \\
\addlinespace[3pt]
\multirow[t]{5}{0.15\textwidth}{Brier score}
& Apparent & 0.0504 (0.0426, 0.0571) & 0.0493 (0.0406, 0.0558) \\
& NEB      & 0.0514 (0.0433, 0.0588) & 0.0513 (0.0435, 0.0588) \\
& OEB      & 0.0514 (0.0440, 0.0587) & 0.0513 (0.0436, 0.0592) \\
& TSB      & 0.0514 (0.0436, 0.0589) & 0.0513 (0.0429, 0.0580) \\
& LSB      & 0.0514 (0.0437, 0.0582) & 0.0513 (0.0425, 0.0578) \\
\addlinespace[3pt]
\multirow[t]{5}{0.15\textwidth}{Log loss}
& Apparent & 0.1872 (0.1617, 0.2086) & 0.1828 (0.1537, 0.2026) \\
& NEB      & 0.1919 (0.1659, 0.2170) & 0.1918 (0.1676, 0.2179) \\
& OEB      & 0.1919 (0.1684, 0.2167) & 0.1918 (0.1675, 0.2190) \\
& TSB      & 0.1919 (0.1649, 0.2152) & 0.1918 (0.1633, 0.2115) \\
& LSB      & 0.1919 (0.1663, 0.2132) & 0.1918 (0.1628, 0.2116) \\
\addlinespace[3pt]
\multirow[t]{5}{0.15\textwidth}{Calibration slope}
& Apparent & 1.000 (1.000, 1.000) & 1.000 (1.000, 1.000) \\
& NEB      & 0.949 (0.800, 1.125) & 0.901 (0.755, 1.053) \\
& OEB      & 0.949 (0.812, 1.120) & 0.901 (0.763, 1.053) \\
& TSB      & 0.949 (0.935, 0.972) & 0.901 (0.877, 0.924) \\
& LSB      & 0.949 (0.949, 0.949) & 0.901 (0.901, 0.901) \\
\addlinespace[3pt]
\multirow[t]{5}{0.15\textwidth}{Calibration intercept}
& Apparent & 0.000 (0.000, 0.000) & 0.000 (0.000, 0.000) \\
& NEB      & -0.111 (-0.506, 0.336) & -0.210 (-0.593, 0.215) \\
& OEB      & -0.111 (-0.470, 0.297) & -0.210 (-0.549, 0.181) \\
& TSB      & -0.111 (-0.146, -0.050) & -0.210 (-0.274, -0.141) \\
& LSB      & -0.111 (-0.111, -0.111) & -0.210 (-0.210, -0.210) \\
\addlinespace[3pt]
\multirow[t]{5}{0.15\textwidth}{Calibration-in-the-large}
& Apparent & 0.000 (0.000, 0.000) & 0.000 (0.000, 0.000) \\
& NEB      & 0.000 (-0.199, 0.190) & 0.000 (-0.205, 0.202) \\
& OEB      & 0.000 (-0.188, 0.203) & 0.000 (-0.195, 0.198) \\
& TSB      & 0.000 (-0.016, 0.023) & 0.000 (-0.017, 0.024) \\
& LSB      & 0.000 (0.000, 0.000) & 0.000 (0.000, 0.000) \\

\bottomrule
\multicolumn{4}{p{0.93\textwidth}}{\footnotesize\textit{Note.} Values are point estimates with 95\% CIs. NEB and TSB use nested bootstrap resampling; OEB and LSB are their corresponding one-level approximations. Bootstrap CIs used 1000 outer samples and, for the nested methods, 100 inner samples within each outer sample. Apparent and DeLong intervals do not apply optimism correction; the DeLong CI is reported only for AUC. \par\noindent\textit{Abbreviations:} NEB, nested error bootstrap; OEB, optimism error bootstrap; TSB, two-stage bootstrap; LSB, location-shifted bootstrap; AUC, area under the receiver operating characteristic curve; CI, confidence interval.}\\
\end{longtable}
\endgroup

\begin{landscape}
\footnotesize
\setlength{\tabcolsep}{2pt}
\renewcommand{\arraystretch}{1.02}
\begin{longtable}{@{}>{\raggedright\arraybackslash}p{0.17\linewidth} >{\raggedright\arraybackslash}p{0.26\linewidth} *{7}{>{\raggedleft\arraybackslash}p{0.065\linewidth}}@{}}
\caption{Approximation of NEB by OEB in the baseline setting of Simulation 2 (synthetic settings).}
\label{tab:neb-oeb-approximation}\\
\toprule
Performance measure & Diagnostic & \multicolumn{7}{c}{EPV} \\
\cmidrule(lr){3-9}
 & & 1 & 3 & 5 & 7 & 10 & 20 & 40 \\
\midrule
\endfirsthead
\multicolumn{9}{l}{\textit{Table \thetable\ continued}}\\
\toprule
Performance measure & Diagnostic & \multicolumn{7}{c}{EPV} \\
\cmidrule(lr){3-9}
 & & 1 & 3 & 5 & 7 & 10 & 20 & 40 \\
\midrule
\endhead
\bottomrule
\endfoot
\endlastfoot

AUC
& OEB/NEB width
& 0.774 & 0.822 & 0.842 & 0.869 & 0.898 & 0.935 & 0.965 \\
& $\sqrt n\,\operatorname{RMSE}_{\star}(\Delta\Omega_H)/\operatorname{SD}_{\star}(R_{\mathrm{FM},n}^\star)$
& 0.729 & 0.324 & 0.247 & 0.202 & 0.169 & 0.125 & 0.111 \\
& $\sqrt n\,\operatorname{RMSE}_{\star}(\Delta\Omega_H)$
& 0.626 & 0.259 & 0.192 & 0.155 & 0.129 & 0.0921 & 0.0812 \\
\addlinespace[2pt]

Brier score
& OEB/NEB width
& 0.777 & 0.959 & 0.968 & 0.972 & 0.977 & 0.982 & 0.990 \\
& $\sqrt n\,\operatorname{RMSE}_{\star}(\Delta\Omega_H)/\operatorname{SD}_{\star}(R_{\mathrm{FM},n}^\star)$
& 0.522 & 0.127 & 0.112 & 0.107 & 0.105 & 0.103 & 0.104 \\
& $\sqrt n\,\operatorname{RMSE}_{\star}(\Delta\Omega_H)$
& 0.192 & 0.0290 & 0.0249 & 0.0237 & 0.0230 & 0.0226 & 0.0226 \\
\addlinespace[2pt]

Log loss
& OEB/NEB width
& 0.800 & 1.14 & 1.02 & 1.00 & 0.997 & 0.991 & 0.996 \\
& $\sqrt n\,\operatorname{RMSE}_{\star}(\Delta\Omega_H)/\operatorname{SD}_{\star}(R_{\mathrm{FM},n}^\star)$
& 0.446 & 0.241 & 0.121 & 0.106 & 0.104 & 0.104 & 0.104 \\
& $\sqrt n\,\operatorname{RMSE}_{\star}(\Delta\Omega_H)$
& 2.15 & 0.220 & 0.0848 & 0.0712 & 0.0681 & 0.0657 & 0.0649 \\
\addlinespace[2pt]

Calibration slope
& OEB/NEB width
& 0.673 & 0.902 & 0.879 & 0.882 & 0.899 & 0.937 & 0.960 \\
& $\sqrt n\,\operatorname{RMSE}_{\star}(\Delta\Omega_H)/\operatorname{SD}_{\star}(R_{\mathrm{FM},n}^\star)$
& 0.811 & 0.350 & 0.238 & 0.195 & 0.167 & 0.124 & 0.108 \\
& $\sqrt n\,\operatorname{RMSE}_{\star}(\Delta\Omega_H)$
& 25.1 & 0.979 & 0.744 & 0.647 & 0.577 & 0.434 & 0.383 \\
\addlinespace[2pt]

Calibration intercept
& OEB/NEB width
& 0.152 & 0.732 & 0.789 & 0.815 & 0.859 & 0.914 & 0.951 \\
& $\sqrt n\,\operatorname{RMSE}_{\star}(\Delta\Omega_H)/\operatorname{SD}_{\star}(R_{\mathrm{FM},n}^\star)$
& 0.952 & 0.381 & 0.280 & 0.228 & 0.188 & 0.131 & 0.110 \\
& $\sqrt n\,\operatorname{RMSE}_{\star}(\Delta\Omega_H)$
& 300 & 2.38 & 1.76 & 1.48 & 1.26 & 0.874 & 0.735 \\
\addlinespace[2pt]

Calibration-in-the-large
& OEB/NEB width
& 1.01 & 1.03 & 1.00 & 1.00 & 1.00 & 0.993 & 0.998 \\
& $\sqrt n\,\operatorname{RMSE}_{\star}(\Delta\Omega_H)/\operatorname{SD}_{\star}(R_{\mathrm{FM},n}^\star)$
& 0.264 & 0.193 & 0.110 & 0.108 & 0.106 & 0.105 & 0.105 \\
& $\sqrt n\,\operatorname{RMSE}_{\star}(\Delta\Omega_H)$
& 9.73 & 0.827 & 0.392 & 0.368 & 0.358 & 0.346 & 0.341 \\

\bottomrule
\multicolumn{9}{p{0.96\linewidth}}{\footnotesize\textit{Note.} $\Delta\Omega_{H,b}=\Omega_{H,n}(P_n^{\star b})-\Omega_{H,n}(P_n)$ and $R_{\mathrm{FM},n}^\star=\sqrt n\{C_{H,n}(P_n^\star)-\Psi(P_n^\star,P_n)\}$.  The baseline setting used eight independent Bernoulli$(0.5)$ predictors, population AUC 0.75, and an event fraction of 0.125. The OEB/NEB width ratio is the median interval width for OEB divided by the median interval width for NEB, with each median calculated among successful replicates for the corresponding method within each scenario and performance measure. The root mean squared error (RMSE) and standard deviation (SD), denoted by $\operatorname{RMSE}_\star$ and $\operatorname{SD}_\star$, respectively, are calculated over the outer bootstrap distribution. The RMSE diagnostics are arithmetic means of replicate-specific values among successful NEB replicates
with finite diagnostic values; the standardized diagnostic is the mean of replicate-specific ratios. These summaries are sensitive to extreme values, particularly at EPV 1.
\par\noindent\textit{Abbreviations:} AUC, area under the receiver operating characteristic curve; EPV, events per variable; NEB, nested error bootstrap; OEB, optimism error bootstrap.}\\
\end{longtable}
\end{landscape}

\clearpage

\section*{Figure legends}

\refstepcounter{figure}\label{fig:gusto-coverage}
\noindent\textbf{Figure \thefigure. Coverage for $\theta_{\mathrm{FM},n}$ in Simulation 1 (GUSTO-I-based settings).} Empirical coverage of 95\% confidence intervals (CIs) is shown across events per variable (EPV) 1, 3, 5, 7, 10, 20, and 40. For each method, coverage was evaluated among replicates in which a CI was successfully constructed and the target value was available. Columns represent the settings with 8 predictors and event fraction 0.125, 8 predictors and event fraction 0.0625, and 17 predictors and event fraction 0.125. Rows show area under the receiver operating characteristic curve (AUC), Brier score, log loss, calibration slope, calibration intercept, and calibration-in-the-large (CITL). Curves show Apparent, nested error bootstrap (NEB), optimism error bootstrap (OEB), basic two-stage bootstrap (bTSB), basic location-shifted bootstrap (bLSB), two-stage bootstrap (TSB), and location-shifted bootstrap (LSB); the dashed horizontal line denotes 0.95.
\par\noindent\textbf{Alt text:}
Six rows and three columns of line plots compare coverage as events per variable increase. Differences among bootstrap methods narrow for the area under the receiver operating characteristic curve, Brier score, and log loss. For calibration, nested error and optimism error bootstrap coverage is close to ninety-five percent at higher events per variable. Both two-stage bootstrap versions remain well below this level, while the apparent method and both location-shifted bootstrap versions have coverage near zero.

\medskip

\refstepcounter{figure}\label{fig:synthetic-coverage}
\noindent\textbf{Figure \thefigure. Coverage for $\theta_{\mathrm{FM},n}$ in Simulation 2 (synthetic settings).} Empirical coverage of 95\% confidence intervals (CIs) is shown across events per variable (EPV) 1, 3, 5, 7, 10, 20, and 40. For each method, coverage was evaluated among replicates in which a CI was successfully constructed and the target value was available. Columns show the baseline setting, the setting with eight continuous predictors, the settings with population area under the receiver operating characteristic curve (AUC) 0.65 and 0.85, the setting with a Pearson correlation of 0.5 among the binary predictors, and the settings with an omitted quadratic or interaction term ($\gamma=1$). Rows show AUC, Brier score, log loss, calibration slope, calibration intercept, and calibration-in-the-large (CITL). All settings used eight predictors and an event fraction of 0.125. The baseline setting used independent Bernoulli$(0.5)$ predictors with population AUC 0.75. The misspecified settings used independent standard normal predictors and retained the linear coefficients from the correctly specified setting with population AUC 0.75. Curves show Apparent, nested error bootstrap (NEB), optimism error bootstrap (OEB), two-stage bootstrap (TSB), and location-shifted bootstrap (LSB); the dashed horizontal line denotes 0.95.
\par\noindent\textbf{Alt text:}
Six rows and seven columns of line plots compare coverage across synthetic settings. At higher events per variable, coverage curves for the bootstrap methods lie close together for the area under the receiver operating characteristic curve, Brier score, and log loss, but remain separated for calibration. For calibration, the nested error and optimism error bootstrap curves are near the nominal level, whereas the two-stage and location-shifted bootstrap curves remain well below it. This pattern also appears in the misspecified settings.

\medskip

\refstepcounter{figure}\label{fig:distributional-diagnostic}
\noindent\textbf{Figure \thefigure. Interval width and diagnostics of the bootstrap distributions in the baseline setting of Simulation 2 (synthetic settings).} The baseline setting used eight independent Bernoulli$(0.5)$ predictors, population area under the receiver operating characteristic curve (AUC) 0.75, and an event fraction of 0.125. Results are shown across events per variable (EPV) 1, 3, 5, 7, 10, 20, and 40. Columns show the median and the 90th and 95th percentiles of confidence interval (CI) width, the mean of $Z_{\mathrm{PIT}}$ (where PIT denotes the probability integral transform), the standard deviation (SD) of $Z_{\mathrm{PIT}}$, and the median standardized Wasserstein distance. CI widths are summarized among replicates with successful interval construction. Rows show AUC, Brier score, log loss, calibration slope, calibration intercept, and calibration-in-the-large (CITL). Curves show Apparent, nested error bootstrap (NEB), optimism error bootstrap (OEB), two-stage bootstrap (TSB), and location-shifted bootstrap (LSB). The Wasserstein distance for every method is standardized by the same standard deviation of the estimation-error root for $\theta_{\mathrm{FM},n}$ within each scenario and performance measure. Boundary symbols indicate values above the plotting range. The annotation in each affected panel gives a lower bound for those values.
\par\noindent\textbf{Alt text:}
Six rows of line plots compare interval widths and distributional diagnostics as events per variable increase. Interval widths generally decrease. For calibration, nested error and optimism error bootstrap intervals remain wider than two-stage and location-shifted bootstrap intervals, while their standardized Wasserstein distances approach zero. Some calibration interval widths at the lowest events per variable exceed the plotting range and are marked at the panel boundaries.

\clearpage
\noindent\textbf{Figure~\ref{fig:gusto-coverage}}

\medskip
{\centering
\makebox[\textwidth][c]{%
\includegraphics[width=\textwidth,height=0.92\textheight,keepaspectratio]{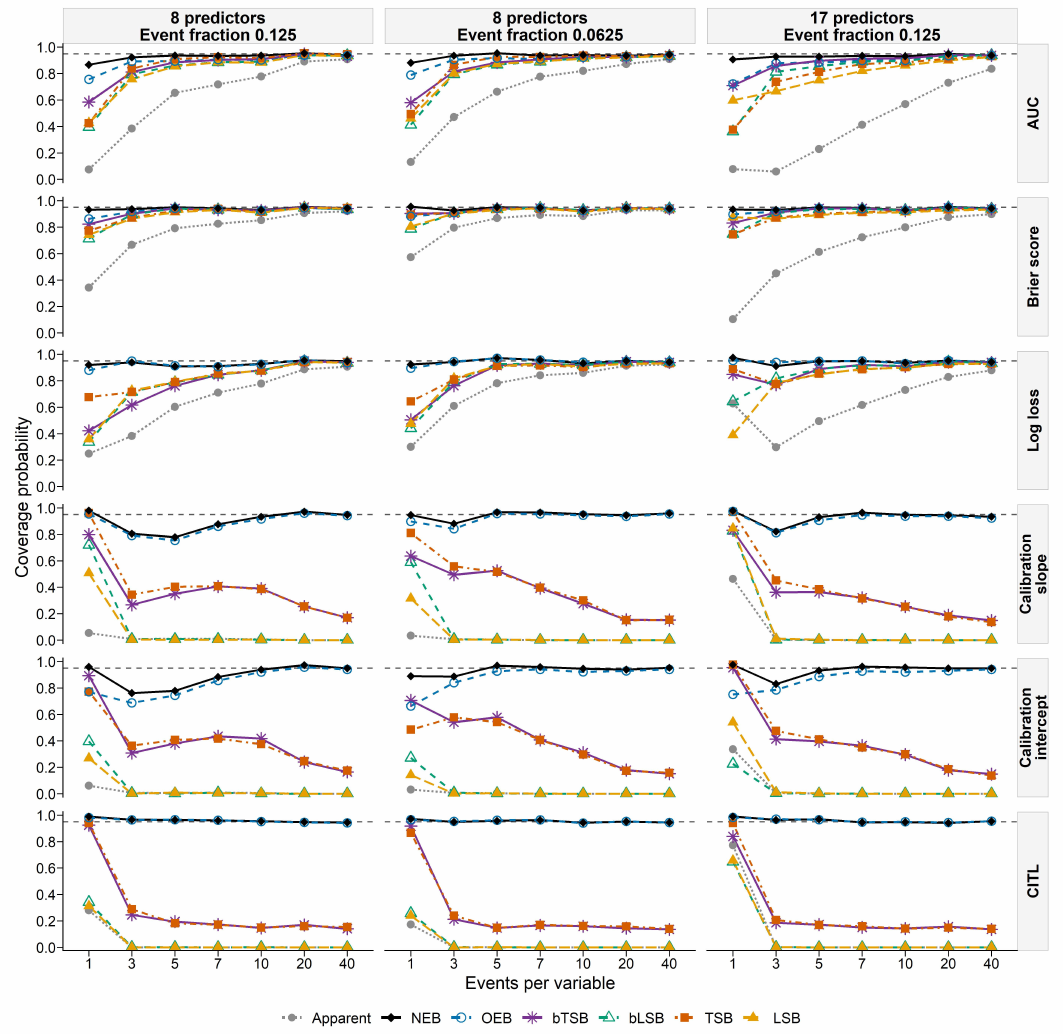}%
}\par}

\clearpage
\noindent\textbf{Figure~\ref{fig:synthetic-coverage}}

\medskip
{\centering
\makebox[\textwidth][c]{%
\includegraphics[width=\textwidth,height=0.92\textheight,keepaspectratio]{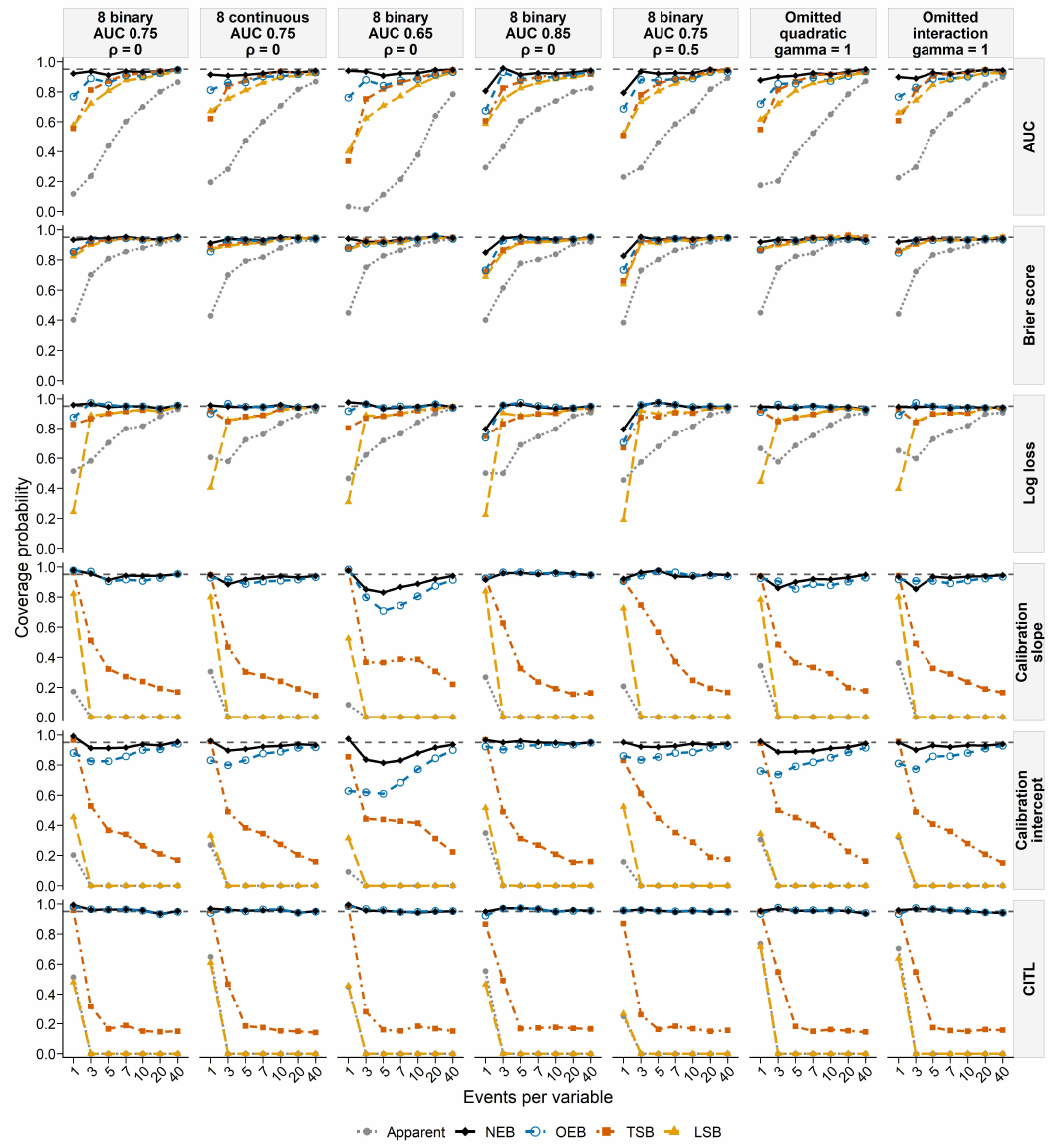}%
}\par}

\clearpage
\noindent\textbf{Figure~\ref{fig:distributional-diagnostic}}

\medskip
{\centering
\makebox[\textwidth][c]{%
\includegraphics[width=\textwidth,height=0.92\textheight,keepaspectratio]{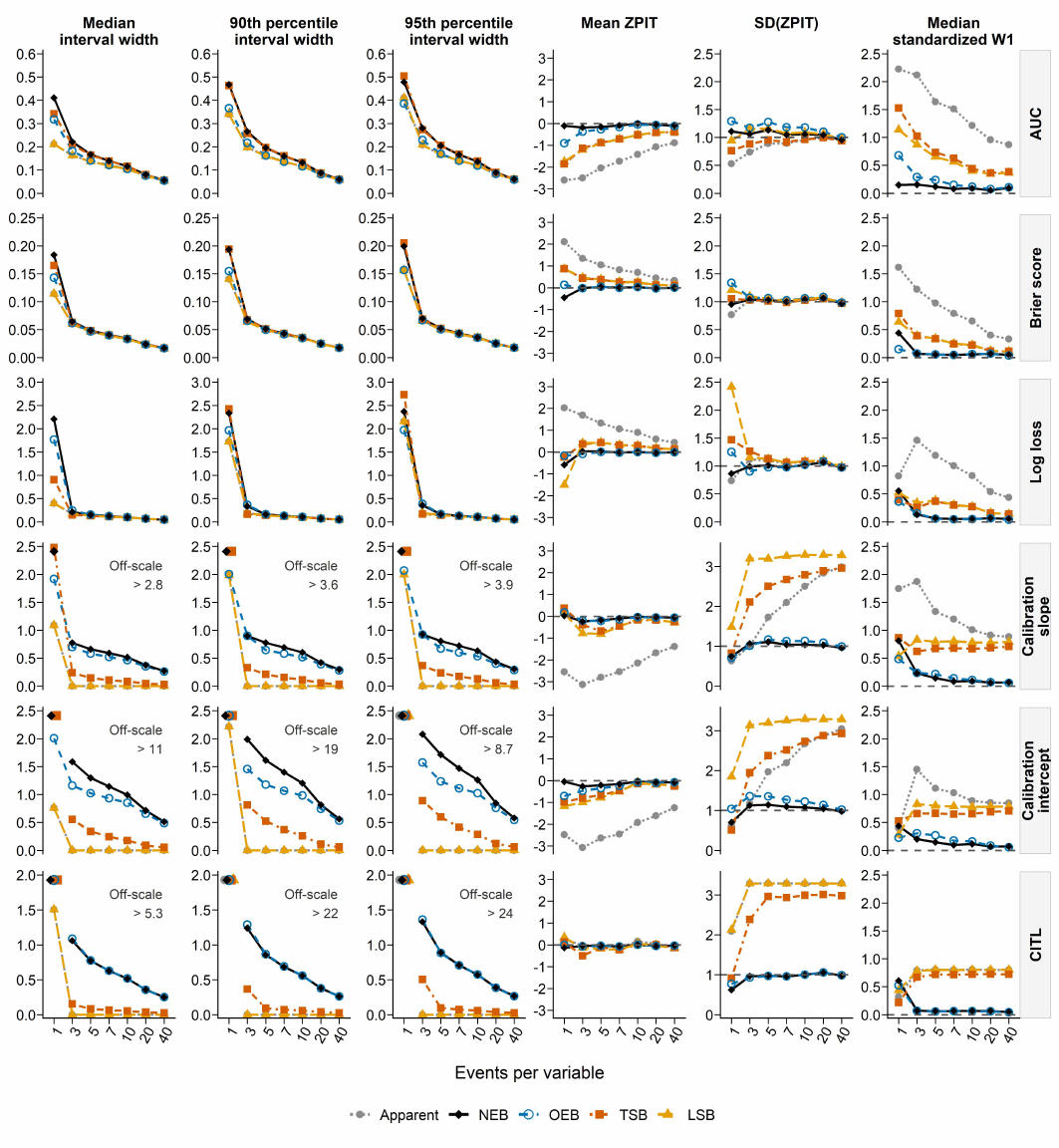}%
}\par}

\medskip
\end{document}


\journaltitle{}\DOI{}\copyrightyear{}\pubyear{}\vol{}\issue{}\access{}\appnotes{}
\firstpage{1}
\title[Supporting Information]{Supporting Information for\\[0.5em]Estimand-specific bootstrap confidence intervals after optimism correction for prediction models}
\author[1,2,$\ast$]{Seiya Takano\ORCID{0000-0001-6990-9923}}
\author[3,4]{Kazuki Nishida}
\author[4]{Ryo Emoto}
\author[1,4]{Shigeyuki Matsui}
\address[1]{Department of Biostatistics and Data Science, School of Public Health, Graduate School of Medicine, Kyoto University, Yoshida-Konoe-cho, Sakyo-ku, Kyoto 606-8501, Japan}
\address[2]{Department of Radiology, Nagoya City University Graduate School of Medical Sciences, 1 Kawasumi, Mizuho-cho, Mizuho-ku, Nagoya, Aichi 467-8601, Japan}
\address[3]{Department of Biostatistics, Nagoya University Graduate School of Medicine, 65 Tsurumai-cho, Showa-ku, Nagoya, Aichi 466-8550, Japan}
\address[4]{Department of Clinical Biostatistics, School of Public Health, Graduate School of Medicine, Kyoto University, Yoshida-Konoe-cho, Sakyo-ku, Kyoto 606-8501, Japan}

\abstract{}
\begingroup\setlength{\parindent}{0pt}\maketitle\endgroup

\section{First-order asymptotics and bootstrap validity for the two estimands}\label{sup:target-roots}

Notation follows the main article. We write $A(P)=\Psi(P,P)$ when the same distribution is used for model development and evaluation. Let $P_n^\star$ denote an empirical bootstrap distribution drawn from $P_n$, and define
\[
\G_n f=\sqrt n(P_n-P_0)f,
\qquad
\G_n^\star f=\sqrt n(P_n^\star-P_n)f.
\]
Throughout, $\rightsquigarrow$ denotes weak convergence, and $\rightsquigarrow_\star$ denotes conditional weak convergence in probability given the observed sample. Unless stated otherwise, the theoretical nested quantities below are defined using the conditional inner-bootstrap expectation, that is, the expectation under the inner-bootstrap distribution given each outer bootstrap sample. In computation, this expectation is approximated by an average over $B_{\mathrm{inner}}$ inner bootstrap samples. The resulting Monte Carlo error is considered separately in Section~\ref{sup:numerical}.

For the generic first-order argument, assume that $\Psi$ admits the expansion
\begin{equation}
\Psi(P,Q)-\Psi(P_0,P_0)
=
(P-P_0)\phi_{\mathrm{tr}}
+
(Q-P_0)\phi_{\mathrm{ev}}
+
r(P,Q),
\label{eqS:local-expansion}
\end{equation}
along the empirical and conditional bootstrap perturbations considered here. We assume $P_0\phi_{\mathrm{tr}}=P_0\phi_{\mathrm{ev}}=0$, square integrability of both influence functions, and remainder terms of order $o_p(n^{-1/2})$ on the relevant empirical paths and $o_{p^\star}(n^{-1/2})$ on the corresponding bootstrap paths. For smooth functionals, such an expansion follows from the functional delta method and its bootstrap analog under appropriate differentiability and empirical-process conditions \citep{vanderVaart1998,vanderVaartWellner2023}.

For the fixed pair $(\phi_{\mathrm{tr}},\phi_{\mathrm{ev}})$, the multivariate central limit theorem and its empirical-bootstrap analog give the corresponding joint Gaussian limits. For the procedure estimand, we additionally require
\[
\E_{P_0^n}|r(P_n^\dagger,P_0)|=o(n^{-1/2}).
\]
Assume also
\begin{equation}
C_{H,n}(P_n)-A(P_n)=o_p(n^{-1/2}),
\qquad
C_{\mathrm{proc},n}(P_n)-C_{H,n}(P_n)=o_p(n^{-1/2}),
\label{eqS:center-conditions}
\end{equation}
with the corresponding conditional bootstrap relations. Section~\ref{app:center-orders} gives sufficient conditions for Brier score, log loss, and the calibration measures. For AUC, these relations are imposed separately, with additional conditions given in Section~\ref{app:auc-scope}.

Applying Equation~\eqref{eqS:local-expansion} gives
\[
\begin{aligned}
A(P_n)
=
\Psi(P_n,P_n)
&=
\Psi(P_0,P_0)
+
(P_n-P_0)(\phi_{\mathrm{tr}}+\phi_{\mathrm{ev}})
+
o_p(n^{-1/2}),\\
\theta_{\mathrm{FM},n}
=
\Psi(P_n,P_0)
&=
\Psi(P_0,P_0)
+
(P_n-P_0)\phi_{\mathrm{tr}}
+
o_p(n^{-1/2}),\\
\theta_{\mathrm{proc},n}
=
\E_{P_0^n}\{\Psi(P_n^\dagger,P_0)\}
&=
\Psi(P_0,P_0)
+
o(n^{-1/2}).
\end{aligned}
\]
The last relation follows because the first-order perturbation from $P_n^\dagger$ has mean zero.

Combining these expansions with Equation~\eqref{eqS:center-conditions} yields
\begin{equation}
\begin{aligned}
R_{\mathrm{FM},n}
:=
\sqrt n\{C_{H,n}(P_n)-\theta_{\mathrm{FM},n}\}
&=
\G_n\phi_{\mathrm{ev}}+o_p(1),\\
R_{\mathrm{proc},n}
:=
\sqrt n\{C_{\mathrm{proc},n}(P_n)-\theta_{\mathrm{proc},n}\}
&=
\G_n(\phi_{\mathrm{tr}}+\phi_{\mathrm{ev}})+o_p(1).
\end{aligned}
\label{eqS:estimand-error-roots}
\end{equation}
By Equation~\eqref{eqS:center-conditions}, the second relation also holds with $C_{H,n}(P_n)$ in place of $C_{\mathrm{proc},n}(P_n)$. Thus, the training contribution cancels from the first-order error for $\theta_{\mathrm{FM},n}$, whereas the error for $\theta_{\mathrm{proc},n}$ retains $\phi_{\mathrm{tr}}+\phi_{\mathrm{ev}}$. For particular performance measures, this distinction can disappear. For log loss, $\phi_{\mathrm{tr}}=0$ when logistic regression is fitted by maximum likelihood without penalization, because the population first-order condition gives $P_0\dot m_{\beta_0}=0$. Here, $m_\beta$ is the loss as a function of the observation, $\dot m_\beta$ is its gradient with respect to $\beta$, and $\beta_0=\beta(P_0)$. For Brier score, the same cancellation holds under correct logistic specification but need not hold under misspecification.

The empirical bootstrap replaces the population distribution $P_0$ by $P_n$ and the observed empirical distribution $P_n$ by $P_n^\star$. Applying the same argument conditionally on the observed sample gives
\begin{equation}
\begin{aligned}
R_{\mathrm{FM},n}^\star
&=
\G_n^\star\phi_{\mathrm{ev}}+o_{p^\star}(1),\\
R_{\mathrm{proc},n}^\star
&=
\G_n^\star(\phi_{\mathrm{tr}}+\phi_{\mathrm{ev}})+o_{p^\star}(1).
\end{aligned}
\label{eqS:bootstrap-reference-roots}
\end{equation}
Let $(G_{\mathrm{tr}},G_{\mathrm{ev}})$ denote the joint Gaussian limit of $(\G_n\phi_{\mathrm{tr}},\G_n\phi_{\mathrm{ev}})$. Then
\[
R_{\mathrm{FM},n}\rightsquigarrow G_{\mathrm{ev}},
\qquad
R_{\mathrm{proc},n}\rightsquigarrow G_{\mathrm{tr}}+G_{\mathrm{ev}},
\]
and the bootstrap roots have the same two limits conditionally. Thus, the two reference roots are bootstrap consistent for the sampling distributions of the estimation errors for their respective estimands. 

\section{Conditions for the one-level approximations}\label{app:fixed-correction}

\subsection{Asymptotic orders of the estimators and optimism correction}\label{app:center-orders}

Section~\ref{sup:target-roots} requires the two differences in Equation~\eqref{eqS:center-conditions} to be $o_p(n^{-1/2})$. For a distribution $P$, let $P_n^\star$ denote the empirical distribution of a size-$n$ sample from $P$, and let $\E_P^\star$ denote expectation over this resampling with $P$ held fixed. From the definitions,
\begin{align}
C_{\mathrm{proc},n}(P)-C_{H,n}(P)
&=
\E_P^\star\{A(P_n^\star)-A(P)\},
\label{eqS:center-gap-identity}\\
C_{H,n}(P)-A(P)
&=
\E_P^\star\{\Psi(P_n^\star,P)-\Psi(P,P)\}
-
\E_P^\star\{A(P_n^\star)-A(P)\}.
\label{eqS:harrell-shift-identity}
\end{align}

For Brier score and log loss, suppose that, locally uniformly over the relevant empirical distributions,
\[
\begin{aligned}
A(P_n^\star)-A(P)
&=
\dot A_P(P_n^\star-P)+R_{A,n}^\star(P),\\
\Psi(P_n^\star,P)-\Psi(P,P)
&=
\dot\Psi_{\mathrm{tr},P}(P_n^\star-P)+R_{\Psi,n}^\star(P),
\end{aligned}
\]
where the first terms are linear and
\[
\E_P^\star|R_{A,n}^\star(P)|
+
\E_P^\star|R_{\Psi,n}^\star(P)|
=
O(n^{-1}).
\]
Since $\E_P^\star(P_n^\star-P)=0$,
\[
\begin{aligned}
\E_P^\star\{A(P_n^\star)-A(P)\}
&=
\E_P^\star R_{A,n}^\star(P),\\
\E_P^\star\{\Psi(P_n^\star,P)-\Psi(P,P)\}
&=
\E_P^\star R_{\Psi,n}^\star(P).
\end{aligned}
\]
Therefore,
\begin{equation}
C_{H,n}(P_n)-A(P_n)=O_p(n^{-1}),
\qquad
C_{\mathrm{proc},n}(P_n)-C_{H,n}(P_n)=O_p(n^{-1}),
\label{eqS:center-orders-smooth}
\end{equation}
with the same orders conditionally under the bootstrap. These rates imply Equation~\eqref{eqS:center-conditions}.

For the calibration measures, Appendix~A of the main article gives
\[
A^{\mathrm{cal}}(P)=c_0,
\qquad
C_{H,n}^{\mathrm{cal}}(P)=C_{\mathrm{proc},n}^{\mathrm{cal}}(P)
\]
whenever the required fitting and calibration solutions are finite and locally unique. Hence,
\[
C_{\mathrm{proc},n}^{\mathrm{cal}}(P)-C_{H,n}^{\mathrm{cal}}(P)=0.
\]
Suppose further, locally uniformly over the relevant empirical distributions, that
\[
\Psi_{\mathrm{cal}}(P_n^\star,P)-\Psi_{\mathrm{cal}}(P,P)
=
\dot\Psi_{\mathrm{tr},P}^{\mathrm{cal}}(P_n^\star-P)
+
R_{\mathrm{cal},n}^\star(P),
\]
where the first term is linear and
\[
\E_P^\star|R_{\mathrm{cal},n}^\star(P)|
=
o(n^{-1/2}).
\]
Then
\[
\begin{aligned}
C_{H,n}^{\mathrm{cal}}(P)-A^{\mathrm{cal}}(P)
&=
\E_P^\star
\{\Psi_{\mathrm{cal}}(P_n^\star,P)-\Psi_{\mathrm{cal}}(P,P)\}\\
&=
\E_P^\star R_{\mathrm{cal},n}^\star(P)
=
o(n^{-1/2}),
\end{aligned}
\]
because $\E_P^\star(P_n^\star-P)=0$. Thus,
\[
C_{H,n}^{\mathrm{cal}}(P_n)-A^{\mathrm{cal}}(P_n)
=
o_p(n^{-1/2}),
\]
with the corresponding conditional order. Together with the exact equality above, this verifies Equation~\eqref{eqS:center-conditions} for the calibration measures.

For AUC, the U-statistic expansion for a fixed prediction score in Section~\ref{app:auc-scope} does not by itself imply these orders. Equation~\eqref{eqS:center-conditions} is therefore imposed separately for AUC.

We next derive the stronger order required when Harrell's bias correction is held fixed across outer bootstrap samples. For a smooth mean-type performance measure, suppose that the fitted prediction rule $\eta_P$ is indexed by a finite-dimensional parameter $\beta(P)$ satisfying $P\psi_{\beta(P)}=0$. Then $\Psi(P,Q)=Qm_P$, where $m_P=m_{\beta(P)}$. Let $J(P)=P\dot\psi_{\beta(P)}$, where $\psi_\beta$ is the estimating function for $\beta$ and the dot denotes differentiation with respect to $\beta$. Let $P_W$ be the empirical distribution of a sample of $n$ independent observations from $P$, and write $\E_W$ for expectation over this sample with $P$ held fixed. Suppose
\[
\begin{aligned}
\beta(P_W)-\beta(P)
&=
-J(P)^{-1}(P_W-P)\psi_{\beta(P)}
+
R_{\beta,W},\\
m_{P_W}
&=
m_P
+
\dot m_{\beta(P)}^{\mathsf T}
\{\beta(P_W)-\beta(P)\}
+
R_{m,W},
\end{aligned}
\]
locally uniformly, with remainder terms contributing $r_{H,n}(P)=o(n^{-1})$ to the bootstrap estimate of optimism.

Since
\[
\Omega_{H,n}(P)
=
\E_W\{(P_W-P)m_{P_W}\},
\]
and $\E_W(P_W-P)f=0$ for fixed $f$,
\[
\begin{aligned}
\Omega_{H,n}(P)
&=
-\E_W
\left[
\{(P_W-P)\dot m_{\beta(P)}\}^{\mathsf T}
J(P)^{-1}
\{(P_W-P)\psi_{\beta(P)}\}
\right]
+
r_{H,n}(P).
\end{aligned}
\]
For a size-$n$ empirical resample from $P$,
\[
\E_W\{(P_W-P)f\,(P_W-P)g\}
=
\frac{1}{n}\{P(fg)-Pf\,Pg\}.
\]
Applying this identity componentwise and using $P\psi_{\beta(P)}=0$ gives
\begin{equation}
\Omega_{H,n}(P)
=
\frac{1}{n}\omega_H(P)+r_{H,n}(P),
\qquad
\omega_H(P)
=
-\operatorname{tr}\!\left[
J(P)^{-1}
P\{\psi_{\beta(P)}\dot m_{\beta(P)}^{\mathsf T}\}
\right].
\label{eqS:optimism-order}
\end{equation}
The leading term is generated by the covariance between perturbations in model fitting and evaluation, as in the covariance-penalty representation of optimism \citep{efron2004}.

For outer bootstrap sample $b$,
\begin{equation}
\Delta\Omega_{H,b}
=
\frac{1}{n}
\{\omega_H(P_n^{\star b})-\omega_H(P_n)\}
+
\{r_{H,n}(P_n^{\star b})-r_{H,n}(P_n)\}.
\label{eqS:optimism-increment-decomposition}
\end{equation}
If
\[
\omega_H(P_n^{\star b})-\omega_H(P_n)
=
O_{p^\star}(n^{-1/2})
\]
and
\[
r_{H,n}(P_n^{\star b})-r_{H,n}(P_n)
=
o_{p^\star}(n^{-3/2}),
\]
then
\begin{equation}
\Delta\Omega_{H,b}
=
O_{p^\star}(n^{-3/2}),
\label{eqS:optimism-increment}
\end{equation}
so that $\sqrt n\,\Delta\Omega_{H,b}=o_{p^\star}(1)$. The analogous condition is required for $\Delta\Omega_{\mathrm{proc},b}$, whereas AUC is treated separately in Section~\ref{app:auc-scope}. 

\subsection{Paired differences and transfer of first-order validity}\label{app:fixed-correction-transfer}

For outer bootstrap sample $b$, define
\[
\Delta\Omega_{H,b}
=
\Omega_{H,n}(P_n^{\star b})-\Omega_{H,n}(P_n),
\qquad
\Delta\Omega_{\mathrm{proc},b}
=
\{A(P_n^{\star b})-C_{\mathrm{proc},n}(P_n^{\star b})\}
-
\{A(P_n)-C_{\mathrm{proc},n}(P_n)\}.
\]

For NEB and OEB,
\[
\begin{aligned}
D_b^{\mathrm{NEB}}
&=
C_{H,n}(P_n)-C_{H,n}(P_n^{\star b})
+\Psi(P_n^{\star b},P_n)\\
&=
D_b^{\mathrm{OEB}}+\Delta\Omega_{H,b}.
\end{aligned}
\]
For TSB and LSB,
\[
\begin{aligned}
D_b^{\mathrm{TSB}}
&=
A(P_n^{\star b})-\Omega_{H,n}(P_n^{\star b}),\\
D_b^{\mathrm{LSB}}
&=
A(P_n^{\star b})-\Omega_{H,n}(P_n),
\end{aligned}
\]
and therefore
\[
D_b^{\mathrm{TSB}}-D_b^{\mathrm{LSB}}
=
-\Delta\Omega_{H,b}.
\]
The corresponding calculation for bTSB and bLSB gives
\[
D_b^{\mathrm{bTSB}}-D_b^{\mathrm{bLSB}}
=
\Delta\Omega_{\mathrm{proc},b}.
\]
Thus,
\begin{equation}
\begin{aligned}
D_b^{\mathrm{NEB}}-D_b^{\mathrm{OEB}}
&=
\Delta\Omega_{H,b},\\
D_b^{\mathrm{TSB}}-D_b^{\mathrm{LSB}}
&=
-\Delta\Omega_{H,b},\\
D_b^{\mathrm{bTSB}}-D_b^{\mathrm{bLSB}}
&=
\Delta\Omega_{\mathrm{proc},b}.
\end{aligned}
\label{eqS:exact-pair-differences}
\end{equation}
The signs differ because TSB and LSB use the percentile bootstrap method, whereas bTSB and bLSB use the basic bootstrap method.

For each pair, let $D_b^{\mathrm{nest}}$ denote the nested member, $D_b^{\mathrm{one}}$ its one-level approximation, and $c_n$ their common centering value. Equation~\eqref{eqS:exact-pair-differences} gives
\[
\sqrt n
\left[
\{D_b^{\mathrm{nest}}-c_n\}
-
\{D_b^{\mathrm{one}}-c_n\}
\right]
=
\pm\sqrt n\,\Delta\Omega_b.
\]
Hence, if $\sqrt n\,\Delta\Omega_b=o_{p^\star}(1)$,
\begin{equation}
d_{\mathrm{BL}}
\left[
\cL_\star\{\sqrt n(D^{\mathrm{nest}}-c_n)\},
\cL_\star\{\sqrt n(D^{\mathrm{one}}-c_n)\}
\right]
\leq
\E_\star\{2\wedge\sqrt n|\Delta\Omega_b|\}
=
o_p(1).
\label{eqS:bl-transfer}
\end{equation}
Here, $\cL_\star$ denotes the conditional law given the observed sample, and $d_{\mathrm{BL}}$ is the bounded Lipschitz distance. Thus, bootstrap consistency of the nested member transfers to its one-level approximation by the triangle inequality. Under continuity and uniqueness of the two nominal limiting quantiles, the same transfer applies to first-order CI coverage.

\section{Additional conditions for AUC}\label{app:auc-scope}

For AUC, let $\eta$ denote a real-valued prediction score used to rank observations, and define
\[
\kappa_\eta(x_1,x_0)
=
\mathbf 1\{\eta(x_1)>\eta(x_0)\}
+
\frac12\mathbf 1\{\eta(x_1)=\eta(x_0)\}.
\]
For the fitted prediction score $\eta_P$, let $\kappa_P=\kappa_{\eta_P}$. The AUC functional is
\begin{equation}
\Psi_{\mathrm{AUC}}(P,Q)
=
\iint
\kappa_P(x_1,x_0)
\,dQ_1^X(x_1)\,dQ_0^X(x_0).
\label{eqS:auc-functional}
\end{equation}
Here, $Q_y^X$ denotes the conditional distribution of $X$ given $Y=y$ under $Q$, for $Q(Y=y)>0$. Let $P_{0,1}^X$ and $P_{0,0}^X$ denote the conditional distributions of $X$ given $Y=1$ and $Y=0$, respectively, and let $\pi_0=P_0(Y=1)\in(0,1)$. For any fixed prediction score $\eta$, define its population AUC under $P_0$ by
\[
A_0(\eta)
=
\iint
\kappa_\eta(x_1,x_0)
\,dP_{0,1}^X(x_1)\,dP_{0,0}^X(x_0).
\]
Write $\eta_0=\eta_{P_0}$ and $\theta_0=A_0(\eta_0)$. For the fixed prediction score $\eta_0$, empirical AUC is a two-sample U-statistic of degree $(1,1)$. Define
\[
V_1(x)
=
\int
\kappa_{\eta_0}(x,x_0)\,dP_{0,0}^X(x_0),
\qquad
V_0(x)
=
\int
\kappa_{\eta_0}(x_1,x)\,dP_{0,1}^X(x_1).
\]
Since $\pi_0\in(0,1)$, the empirical proportions of cases and controls converge in probability to $\pi_0$ and $1-\pi_0$, respectively. The H\'ajek projection
\citep{hoeffding1948,serfling1980,delong1988}
therefore gives
\begin{equation}
\phi_{\mathrm{ev}}^{\mathrm{AUC}}(z)
=
\frac{y}{\pi_0}\{V_1(x)-\theta_0\}
+
\frac{1-y}{1-\pi_0}\{V_0(x)-\theta_0\}.
\label{eqS:auc-evaluation-if}
\end{equation}

The fitted prediction score changes with the development sample. Therefore, the U-statistic expansion for a fixed prediction score alone is insufficient. For two prediction scores $\eta$ and $\widetilde\eta$, define
\begin{equation}
\rho^2(\eta,\widetilde\eta)
=
\iint
\{\kappa_\eta(x_1,x_0)-\kappa_{\widetilde\eta}(x_1,x_0)\}^2
\,dP_{0,1}^X(x_1)\,dP_{0,0}^X(x_0).
\label{eqS:auc-semimetric}
\end{equation}
For
\[
V_{1,\eta}(x)
=
\int
\kappa_\eta(x,x_0)\,dP_{0,0}^X(x_0),
\qquad
V_{0,\eta}(x)
=
\int
\kappa_\eta(x_1,x)\,dP_{0,1}^X(x_1),
\]
the Cauchy--Schwarz inequality gives
\[
\|V_{1,\eta}-V_{1,\widetilde\eta}\|_{L_2(P_{0,1}^X)}
\leq
\rho(\eta,\widetilde\eta),
\qquad
\|V_{0,\eta}-V_{0,\widetilde\eta}\|_{L_2(P_{0,0}^X)}
\leq
\rho(\eta,\widetilde\eta),
\]
and
\[
|A_0(\eta)-A_0(\widetilde\eta)|
\leq
\rho(\eta,\widetilde\eta).
\]
Thus, convergence in $\rho$ implies $L_2$ convergence of the corresponding H\'ajek projections and convergence of the corresponding population AUC.

Assume
\[
\rho(\eta_{P_n},\eta_0)
\overset{p}{\longrightarrow}
0
\]
and, conditionally on the observed sample,
\[
\rho(\eta_{P_n^\star},\eta_{P_n})
=
o_{p^\star}(1).
\]
For the fixed-dimensional logistic model considered here, the ordering kernels are determined by half-spaces in the case--control predictor difference and form a bounded Vapnik--Chervonenkis subgraph class. Under the usual measurability conditions, standard U-process and bootstrap equicontinuity results therefore apply \citep{vanderVaartWellner2023,arconesGine1993,arconesGine1994}.

Ordinary and bootstrap consistency of the logistic regression coefficient estimator, together with
\begin{equation}
(P_{0,1}^X\otimes P_{0,0}^X)
\left\{
\eta_0(X_1)=\eta_0(X_0),\;
X_1\neq X_0
\right\}
=
0,
\label{eqS:auc-tie-condition}
\end{equation}
is sufficient for the two convergence relations in $\rho$. The condition allows ties caused by identical predictor patterns, which remain ties under perturbations of the fitted coefficients. Condition~\eqref{eqS:auc-tie-condition} is not satisfied in the binary-predictor scenarios with equal population coefficients in Simulation 2, because distinct predictor patterns can have identical population scores. We therefore rely on the simulation results to assess the performance of the bootstrap CIs for AUC in these scenarios.

Under the preceding regularity conditions, replacing the limiting prediction score by the fitted prediction score does not affect the first-order limit of the centered evaluation process. Since
\[
\theta_{\mathrm{FM},n}^{\mathrm{AUC}}
=
\Psi_{\mathrm{AUC}}(P_n,P_0),
\]
we have
\begin{align}
\sqrt n
\{\Psi_{\mathrm{AUC}}(P_n,P_n)-\theta_{\mathrm{FM},n}^{\mathrm{AUC}}\}
&=
\G_n\phi_{\mathrm{ev}}^{\mathrm{AUC}}+o_p(1),\\
\sqrt n
\{\Psi_{\mathrm{AUC}}(P_n^\star,P_n^\star)-\Psi_{\mathrm{AUC}}(P_n^\star,P_n)\}
&=
\G_n^\star\phi_{\mathrm{ev}}^{\mathrm{AUC}}+o_{p^\star}(1).
\label{eqS:auc-evaluation-expansions}
\end{align}
The second expansion is the conditional-bootstrap analog of the first under the substitution $(P_n,P_0)\mapsto(P_n^\star,P_n)$. In each difference, the training distribution is fixed and only the evaluation distribution changes. Thus, the first-order root for $\theta_{\mathrm{FM},n}^{\mathrm{AUC}}$ requires only the expansion with respect to the evaluation distribution.

For inference on $\theta_{\mathrm{proc},n}^{\mathrm{AUC}}$, we additionally require a first-order expansion with respect to the training distribution:
\begin{equation}
\sqrt n
\{\Psi_{\mathrm{AUC}}(P_n,P_0)-\Psi_{\mathrm{AUC}}(P_0,P_0)\}
=
\sqrt n\{A_0(\eta_{P_n})-A_0(\eta_0)\}
=
\G_n\phi_{\mathrm{tr}}^{\mathrm{AUC}}+o_p(1),
\label{eqS:auc-training-expansion}
\end{equation}
together with its locally uniform conditional-bootstrap analog. This condition is stronger than convergence in $\rho$ and does not follow from the VC property. For the fixed-dimensional logistic model, one sufficient set of conditions is asymptotic linearity of the coefficient estimator and differentiability of $A_0(\eta_\beta)$ at the population coefficient, which requires suitable regularity of the case--control score margin near zero.

Finally, the exact paired identities in Equation~\eqref{eqS:exact-pair-differences} do not require smoothness of the AUC functional. The transfer from NEB to OEB requires the additional condition
\[
\sqrt n\,\Delta\Omega_{H,b}^{\mathrm{AUC}}
=
o_{p^\star}(1).
\]

\section{Assessment of Monte Carlo error and bootstrap distributions}\label{sup:numerical}

\subsection{Selection and sensitivity of the number of inner bootstrap samples}\label{sup:k-selection}

Before the final simulation, we conducted a pilot in the setting based on the GUSTO-I West region data with 8 predictors and an event fraction of 0.125. The pilot used 150 simulation replicates and 500 outer bootstrap samples per replicate. We examined $B_{\mathrm{inner}}=50,100,200,400,600,800,$ and $1000$ at EPV 20 and 40. The pilot was used only to select the number of inner bootstrap samples.

For simulation replicate $r$, method $\mathcal M$, and candidate $B_{\mathrm{inner}}$, let $\widehat{\operatorname{MCSE}}_{r,b,B_{\mathrm{inner}}}^{\mathcal M}$ denote the estimated Monte Carlo standard error of the inner correction in outer bootstrap sample $b$. Let $D_{r,b,B_{\mathrm{inner}}}^{\mathcal M}$ denote the corresponding bootstrap replicate used to form the CI. Performance-measure indices are suppressed. We defined
\begin{equation}
\rho_{r,B_{\mathrm{inner}}}^{\mathcal M}
=
\frac{
\operatorname{RMS}_{b}
\{\widehat{\operatorname{MCSE}}_{r,b,B_{\mathrm{inner}}}^{\mathcal M}\}
}{
\operatorname{SD}_{b}
\{D_{r,b,B_{\mathrm{inner,max}}}^{\mathcal M}\}
},
\qquad
\rho_{90}^{\mathcal M}(B_{\mathrm{inner}})
=
Q_{0.90,r}
\{\rho_{r,B_{\mathrm{inner}}}^{\mathcal M}\},
\label{eqS:rho}
\end{equation}
where $B_{\mathrm{inner,max}}=1000$, $\operatorname{RMS}_b$ denotes the root mean square over outer bootstrap samples, and $Q_{0.90,r}$ denotes the empirical 90th percentile across simulation replicates. Under an additive, conditionally mean-zero approximation for the inner Monte Carlo error, the standard deviation of the bootstrap distribution is inflated by approximately
\[
\sqrt{1+\rho_{r,B_{\mathrm{inner}}}^2}-1.
\]
Thus, $\rho_{90}<0.14$ corresponds to less than approximately 1\% inflation in the standard deviation for 90\% of simulation replicates.

For confidence limits $L_{r,B_{\mathrm{inner}}}$ and $U_{r,B_{\mathrm{inner}}}$, we also defined
\begin{equation}
D_{r,B_{\mathrm{inner}}}
=
\frac{
\max\{|L_{r,B_{\mathrm{inner}}}-L_{r,B_{\mathrm{inner,max}}}|,
|U_{r,B_{\mathrm{inner}}}-U_{r,B_{\mathrm{inner,max}}}|\}
}{
\{U_{r,B_{\mathrm{inner,max}}}-L_{r,B_{\mathrm{inner,max}}}\}/2
},
\qquad
D_{90}(B_{\mathrm{inner}})
=
Q_{0.90,r}\{D_{r,B_{\mathrm{inner}}}\}.
\label{eqS:D}
\end{equation}
The criterion $D_{90}<0.06$ limits the movement of either confidence limit to 3\% of the reference interval width for approximately 90\% of simulation replicates.

We selected the smallest $B_{\mathrm{inner}}$ satisfying $\rho_{90}<0.14$ and $D_{90}<0.06$ in every eligible cell. Eligible cells comprised NEB for all six performance measures and TSB and bTSB for AUC, Brier score, and log loss at EPV 20 and 40. Coverage and interval width were not used for selection. Table~\ref{tabS:precision} gives the results.

\begin{table}[!htbp]
\centering
\footnotesize
\setlength{\tabcolsep}{5pt}
\renewcommand{\arraystretch}{1.20}
\setlength{\extrarowheight}{1pt}

\caption{Selection of the number of inner bootstrap samples.}
\label{tabS:precision}

\begin{tabularx}{\linewidth}{
@{}
r
r
>{\raggedright\arraybackslash}X
r
>{\raggedright\arraybackslash}X
@{}
}
\toprule
$B_{\mathrm{inner}}$
& Largest $\rho_{90}$
& Cell with largest $\rho_{90}$
& Largest $D_{90}$
& Cell with largest $D_{90}$ \\
\midrule
50
& 0.156
& EPV 20, log loss, TSB
& 0.0613
& EPV 40, calibration slope, NEB \\

100
& 0.110
& EPV 20, log loss, TSB
& 0.0505
& EPV 40, calibration intercept, NEB \\

200
& 0.0779
& EPV 20, log loss, TSB
& 0.0371
& EPV 20, AUC, TSB \\

400
& 0.0550
& EPV 20, log loss, TSB
& 0.0251
& EPV 20, AUC, TSB \\

600
& 0.0449
& EPV 20, log loss, TSB
& 0.0190
& EPV 20, calibration intercept, NEB \\

800
& 0.0389
& EPV 20, log loss, TSB
& 0.0117
& EPV 20, calibration intercept, NEB \\

1000
& 0.0348
& EPV 20, log loss, TSB
& 0
& EPV 20, AUC, NEB \\
\bottomrule
\end{tabularx}

\vspace{5pt}

\begin{minipage}{0.98\linewidth}
\footnotesize
\textit{Note.}
The pilot used the GUSTO-I West region setting with 8 predictors and an event fraction of 0.125, with 150 simulation replicates and 500 outer bootstrap samples per replicate.
Eligible cells included NEB for all six performance measures and TSB and bTSB for AUC, Brier score, and log loss at EPV 20 and 40.
The smallest $B_{\mathrm{inner}}$ satisfying $\rho_{90}<0.14$ and $D_{90}<0.06$ in every eligible cell was 100.
Coverage and interval width were not used for selection.
\end{minipage}

\end{table}

\FloatBarrier

\begin{figure}[p]
\centering
\includegraphics[
  width=\textwidth,
  height=0.84\textheight,
  keepaspectratio
]{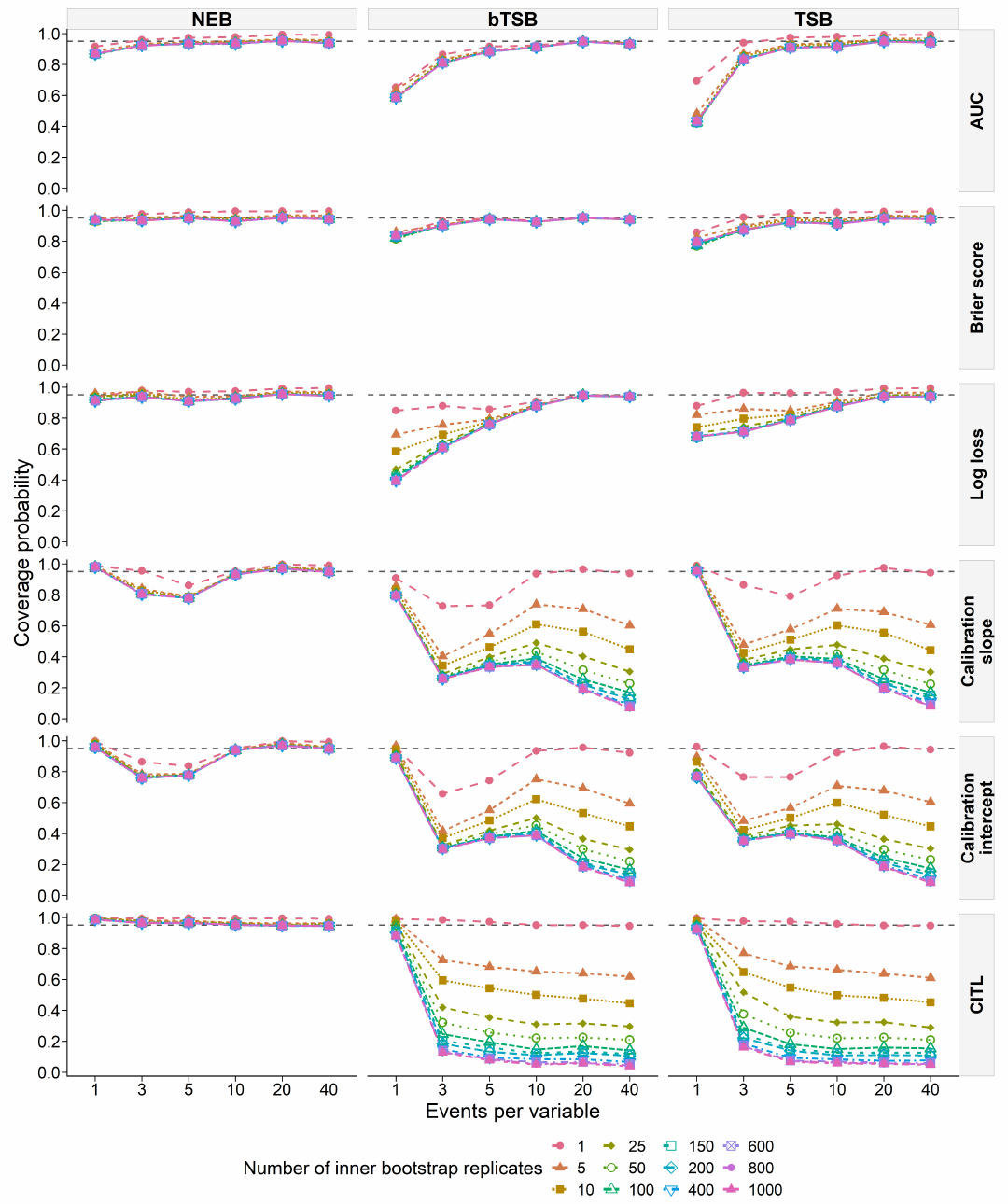}
\vspace{2pt}
\caption{Sensitivity of coverage for $\theta_{\mathrm{FM},n}$ to the number of inner bootstrap samples. Empirical coverage of 95\% confidence intervals (CIs) is shown for nested error bootstrap (NEB), basic two-stage bootstrap (bTSB), and two-stage bootstrap (TSB) in the GUSTO-I-based simulation setting with 8 predictors and an event fraction of 0.125. For each value of $B_{\mathrm{inner}}$, coverage was evaluated among replicates in which a CI was successfully constructed and the target value was available. Columns show the three methods, and rows show area under the receiver operating characteristic curve (AUC), Brier score, log loss, calibration slope, calibration intercept, and calibration-in-the-large (CITL). Results are shown across events per variable (EPV) 1, 3, 5, 10, 20, and 40 for $B_{\mathrm{inner}}=1,5,10,25,50,100,150,200,400,600,800,$ and $1000$. The dashed horizontal line denotes the nominal coverage probability of 0.95.}
\label{figS:coverage}
\end{figure}

\begin{figure}[p]
\centering
\includegraphics[
  width=\textwidth,
  height=0.65\textheight,
  keepaspectratio
]{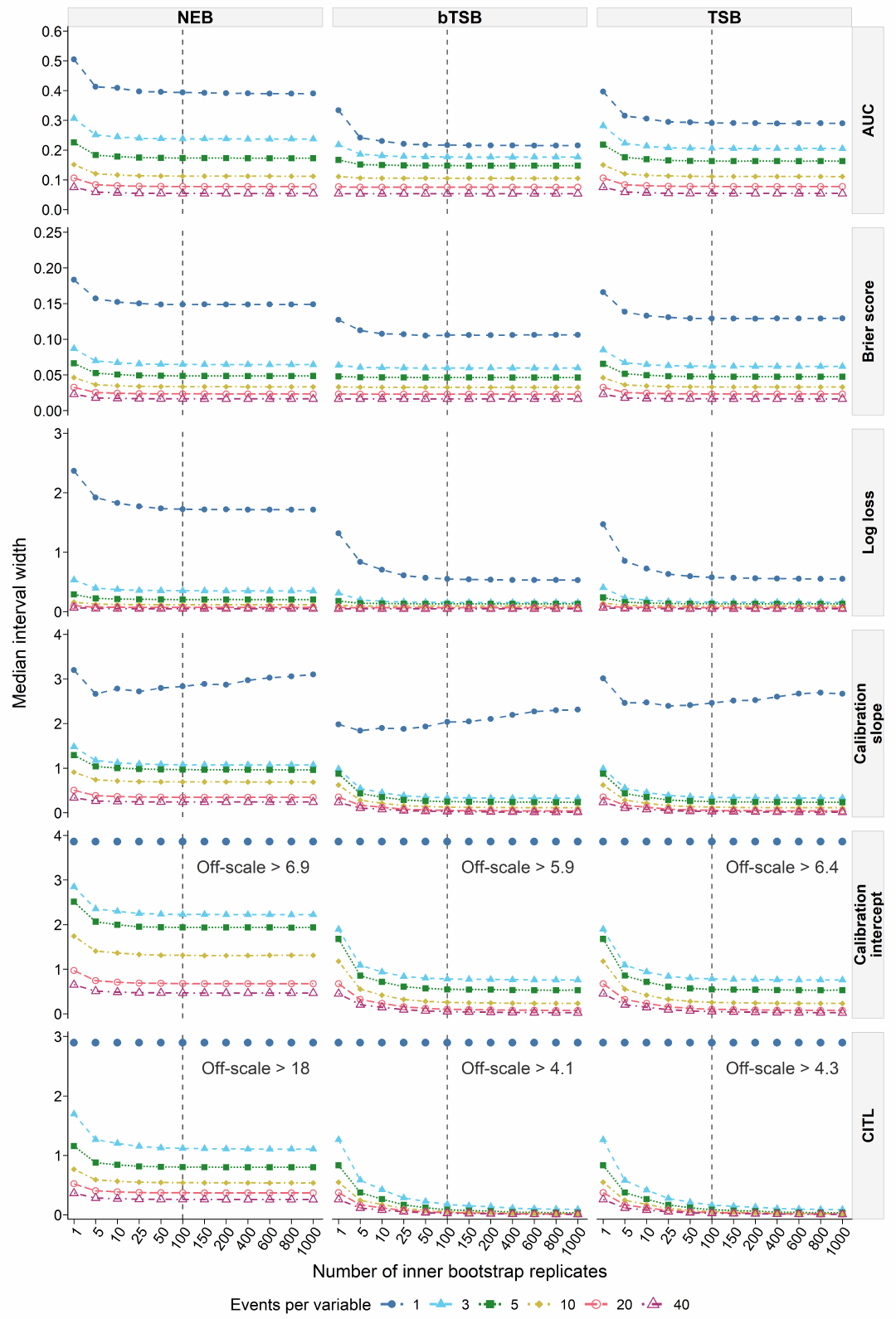}
\vspace{2pt}
\caption{Sensitivity of median interval width to the number of inner bootstrap samples. Median widths of 95\% confidence intervals (CIs) are shown for nested error bootstrap (NEB), basic two-stage bootstrap (bTSB), and two-stage bootstrap (TSB) in the GUSTO-I-based simulation setting with 8 predictors and an event fraction of 0.125. For each value of $B_{\mathrm{inner}}$, the median CI width was calculated among replicates in which a CI was successfully constructed. Columns show the three methods, and rows show area under the receiver operating characteristic curve (AUC), Brier score, log loss, calibration slope, calibration intercept, and calibration-in-the-large (CITL). Curves represent events per variable (EPV) 1, 3, 5, 10, 20, and 40 across values of $B_{\mathrm{inner}}$ from 1 to 1000. The vertical dashed line denotes $B_{\mathrm{inner}}=100$, which was used in the main analysis. Values exceeding the plotting range are truncated at the upper limit, with their actual values shown as annotations.}
\label{figS:width}
\end{figure}

After $B_{\mathrm{inner}}=100$ was fixed from the pilot analysis, all main analyses used this value. The cumulative $B_{\mathrm{inner}}$ results from the final simulations were then used for exploratory sensitivity analyses of coverage and interval width. Supplementary Figures~\ref{figS:coverage} and~\ref{figS:width} show coverage for $\theta_{\mathrm{FM},n}$ and median interval width, respectively, across the evaluated values of $B_{\mathrm{inner}}$. These analyses did not affect the choice of $B_{\mathrm{inner}}$. Median widths changed little beyond $B_{\mathrm{inner}}=100$ for most performance measures and EPV values, whereas large median widths persisted for the calibration measures at low EPV.

For the calibration measures, TSB and bTSB had coverage close to the nominal level when $B_{\mathrm{inner}}=1$, although neither showed a consistent advantage in coverage over NEB or OEB across performance measures. Their intervals became narrower as $B_{\mathrm{inner}}$ increased. This pattern is consistent with the cancellation of first-order variation for inference on $\theta_{\mathrm{proc},n}^{\mathrm{cal}}$ shown in Appendix~A of the main article. Since $\Psi_{\mathrm{cal}}(P,P)=c_0$, the estimator based on Harrell's bias correction within outer bootstrap sample $b$, using a finite number of inner bootstrap samples, is
\[
\widehat C_{H,n,B_{\mathrm{inner}}}^{\mathrm{cal}}(P_n^{\star b})
=
\frac{1}{B_{\mathrm{inner}}}
\sum_{j=1}^{B_{\mathrm{inner}}}
\Psi_{\mathrm{cal}}
\{P_n^{\star\star(b,j)},P_n^{\star b}\}.
\]
Write $\mathbb G_{n,bj}^{\star\star}f=\sqrt n(P_n^{\star\star(b,j)}-P_n^{\star b})f$. Conditional on $P_n^{\star b}$, each inner resample perturbs the training distribution while the evaluation distribution remains fixed. Hence, to first order, $\sqrt n\{\Psi_{\mathrm{cal}}(P_n^{\star\star(b,j)},P_n^{\star b})-c_0\}=\mathbb G_{n,bj}^{\star\star}\phi_{\mathrm{tr}}^{\mathrm{cal}}+o_{p^\star}(1)$. Because $\phi_{\mathrm{tr}}^{\mathrm{cal}}+\phi_{\mathrm{ev}}^{\mathrm{cal}}=0$, averaging over $B_{\mathrm{inner}}$ independent inner resamples gives variance approximately $\Var_{P_0}(\phi_{\mathrm{ev}}^{\mathrm{cal}})/B_{\mathrm{inner}}$ on the $\sqrt n$ scale. In comparison, $\sqrt n(\theta_{\mathrm{FM},n}^{\mathrm{cal}}-c_0)=-\mathbb G_n\phi_{\mathrm{ev}}^{\mathrm{cal}}+o_p(1)$. Thus, when $B_{\mathrm{inner}}=1$, the finite-inner variation of TSB has the same first-order component and variance as the sampling variation of $\theta_{\mathrm{FM},n}^{\mathrm{cal}}$. As $B_{\mathrm{inner}}$ increases, this variation is averaged out and the cancellation for inference on $\theta_{\mathrm{proc},n}^{\mathrm{cal}}$ becomes apparent. The same variance reduction applies to bTSB.

This sensitivity analysis was designed to assess the numerical stability of the nested bootstrap CIs at $B_{\mathrm{inner}}=100$. Evaluation of TSB and bTSB with $B_{\mathrm{inner}}=1$ as practical methods for inference was therefore beyond the scope of this study. The observed pattern nevertheless suggests that characterizing first-order variation through influence functions may be useful for interpreting CI width as well as coverage.

\subsection{Definitions of PIT and Wasserstein distance diagnostics}\label{sup:root}

For method $m$ in simulation replicate $r$, let $D_{rmb}^\star$, $b=1,\ldots,B_{r,m}$, denote the bootstrap replicates, and let $\theta_{r,\mathrm{FM}}$ denote the target $\theta_{\mathrm{FM},n}$. To assess the calibration of each bootstrap distribution for $\theta_{\mathrm{FM},n}$ across simulation replicates, we used the probability integral transform (PIT). The empirical PIT is discrete when based on a finite number of replicates \citep{sailynoja2022}. We used the following mid-rank version:
\begin{equation}
\operatorname{PIT}_{r,m}
=
\frac{\frac12+\sum_{b=1}^{B_{r,m}}\mathbf 1(D_{rmb}^\star<\theta_{r,\mathrm{FM}})+\frac12\sum_{b=1}^{B_{r,m}}\mathbf 1(D_{rmb}^\star=\theta_{r,\mathrm{FM}})}
{B_{r,m}+1},
\qquad
Z_{\mathrm{PIT},r,m}=\Phi^{-1}\{\operatorname{PIT}_{r,m}\}.
\label{eqS:pit}
\end{equation}
This form gives half weight to ties and keeps the PIT in $(0,1)$. If the bootstrap distribution is well calibrated, the PIT is approximately uniform across simulation replicates, and $Z_{\mathrm{PIT}}$ has mean near 0 and standard deviation near 1 \citep{gneitingBalabdaouiRaftery2007}. 

To quantify the distance between each bootstrap distribution and the sampling distribution of the estimation error for $\theta_{\mathrm{FM},n}$, we used the standardized first Wasserstein distance $W_1$ \citep{peyreCuturi2019}. Compared with $d_{\mathrm{BL}}$, $W_1$ retains the magnitude of differences in location, dispersion, and tails on the scale of the root. When first moments are finite, the paired nested and one-level bootstrap distributions satisfy
\begin{equation}
d_{\mathrm{BL}}
\{\cL_\star(D^{\mathrm{nest}}),\cL_\star(D^{\mathrm{one}})\}
\leq
W_1
\{\cL_\star(D^{\mathrm{nest}}),\cL_\star(D^{\mathrm{one}})\}
\leq
\E_\star|\Delta\Omega_b|.
\label{eqS:w1-bound}
\end{equation}

Let $C_{r,m}$ be the reported point estimate, $Q_{r,m}^\star$ the quantile function of $\sqrt n(D_{rmb}^\star-C_{r,m})$, and $\widehat Q_{-r,m,\mathrm{FM}}$ the leave-one-out empirical quantile function of $\sqrt n(\theta_{s,\mathrm{FM}}-C_{s,m})$ over $s\neq r$. The leave-one-out construction avoids including replicate $r$ in its own reference distribution. Using the common midpoint grid $p_j=(j-1/2)/J$, $j=1,\ldots,J$, with $J=201$, we calculated
\begin{equation}
\widehat W_{1,r,m}
=
\frac{1}{J}\sum_{j=1}^J
\left|Q_{r,m}^\star(p_j)-\widehat Q_{-r,m,\mathrm{FM}}(p_j)\right|.
\label{eqS:w1}
\end{equation}

We standardized the distance by
\begin{equation}
\widehat W_{1,r,m}^{\mathrm{std}}
=
\frac{\widehat W_{1,r,m}}{\widehat\sigma_{\mathrm{ref},\mathrm{FM}}},
\qquad
\widehat\sigma_{\mathrm{ref},\mathrm{FM}}
=
\operatorname{SD}_r\left[
\sqrt n\{C_{H,n}(P_{n,r})-\theta_{\mathrm{FM},n,r}\}
\right].
\label{eqS:w1-standardized}
\end{equation}
The denominator was common to all methods within each scenario and performance measure so that differences in scale between methods were retained. We summarized standardized $W_1$ by its median across simulation replicates.

\section{Additional details of the GUSTO-I data and synthetic simulations}
\label{sup:data-details}

\subsection{GUSTO-I West region data}
\label{sup:real-data}

Following \citet{noma2021}, we used the prespecified sets of 8 and 17 predictors from the GUSTO-I West region data \citep{steyerberg2001}. The 8 predictors were age greater than 65 years (A65), sex (SEX), diabetes (DIA), hypotension (HYP), tachycardia or high heart-rate status (HRT), high-risk status (HIG), shock (SHO), and no relief of chest pain (TTR). The 17-predictor set additionally included previous myocardial infarction (PMI), height (HEI), weight (WEI), hypertension (HTN), smoking (SMK), hypercholesterolemia (LIP), previous angina (PAN), family history of myocardial infarction (FAM), and ST elevation in more than four leads (ST4). Binary variables retained 0/1 coding. Smoking was represented by two indicator variables, $\mathrm{SMK1}=\mathbf 1(\mathrm{SMK}=1)$ and $\mathrm{SMK2}=\mathbf 1(\mathrm{SMK}=2)$, with category 3 as the reference. For EPV, smoking was counted as one predictor before indicator coding. Both analyses included 2188 complete cases and 135 deaths within 30 days. The prespecified 8- and 17-predictor sets corresponded to 16.9 and 7.94 events per variable (EPV), respectively.

\subsection{Data-generating mechanisms for Simulation 2}\label{sup:synthetic-dgp}

The synthetic settings were based in part on simulation factors considered by van Smeden et al., including population AUC, predictor distribution, predictor correlation, and equal predictor effects \citep{vanSmeden2019}. For the correctly specified settings, we used population AUC values of 0.65, 0.75, and 0.85, Bernoulli$(0.5)$ predictors, and standard normal predictors.

For the correctly specified synthetic settings, outcomes were generated from
\[
\operatorname{logit}\{P(Y=1\mid X)\}
=
\beta_0(s)+s\,b^{\mathsf T}X,
\]
where $b$ specifies the relative predictor effects, $s\geq0$ is a common scale, and $\beta_0(s)$ is the intercept chosen for a given $s$. For all settings reported here, $b=(1,\ldots,1)^{\mathsf T}$.

For each correctly specified scenario, $s$ was selected using 200,000 predictor vectors. For each candidate $s$, $\beta_0(s)$ was chosen so that the mean event probability was 0.125. Population AUC was approximated on the same sample using $p_i(s)=\expit\{\beta_0(s)+s\,b^{\mathsf T}X_i\}$ as the case weight, $1-p_i(s)$ as the control weight, and $s\,b^{\mathsf T}X_i$ as the score, with half weight for ties. A one-dimensional root search selected $s$ so that this AUC matched the prespecified value. After $s$ was fixed, the intercept used to generate the simulation data was recalculated using an independent sample of 100,000 predictor vectors to obtain an event fraction of 0.125. These samples were separate from those used to approximate $\theta_{\mathrm{FM},n}$ in each simulation replicate.

The correlated binary setting used eight Bernoulli$(0.5)$ predictors with pairwise population Pearson correlation 0.5. They were generated by thresholding an equicorrelated standard multivariate normal vector at zero. The latent correlation was set to
\[
\rho_{\mathrm{latent}}
=
\sin\left(\frac{\pi}{2}\times0.5\right)
=
\frac{1}{\sqrt{2}},
\]
which gives Pearson correlation 0.5 after dichotomization.

For the two misspecified settings, $X_j\overset{\mathrm{iid}}{\sim}N(0,1)$ and
\[
\operatorname{logit}\{P(Y=1\mid X)\}
=
\alpha_{\gamma,h}+\beta^{\mathsf T}X+\gamma h(X),
\]
where $h(X)=(X_1^2-1)/\sqrt{2}$ or $X_1X_2$ and $\gamma=1$. The linear coefficients were retained from the correctly specified continuous-predictor setting with target population AUC 0.75. For each choice of $ h(X)$, we adjusted the intercept to target an event fraction of 0.125. The fitted model included only the eight linear terms. 

\section{ABCLOC and its relation to OEB}\label{sup:abcloc}

Harrell described asymmetric bootstrap confidence limits for overfitting-corrected model performance metrics (ABCLOC) in an online report \citep{harrell2025}. We examine its relation to OEB and its coverage for $\theta_{\mathrm{FM},n}$.

\subsection{Definition and relation to OEB}\label{sup:abcloc-definition}

Using the notation of the main article, write
\begin{equation}
 A_n=\Psi(P_n,P_n),\qquad
 A_b^\star=\Psi(P_n^{\star b},P_n^{\star b}),\qquad
 T_b^\star=\Psi(P_n^{\star b},P_n).
 \label{eqS:abcloc-inputs}
\end{equation}
Thus, $A_b^\star$ and $T_b^\star$ evaluate the same bootstrap-fitted model in the bootstrap and original samples, respectively. For $B$ bootstrap samples, let
\begin{equation}
 O_b^\star=A_b^\star-T_b^\star,\qquad
 \bar O^\star=B^{-1}\sum_{b=1}^B O_b^\star,\qquad
 \widehat C_{H,n}(P_n)=A_n-\bar O^\star.
 \label{eqS:abcloc-center}
\end{equation}
Here, $\widehat C_{H,n}(P_n)$ is the Monte Carlo estimate of Harrell's optimism-corrected estimator $C_{H,n}(P_n)$ \citep{harrell1996,efron1983}, which NEB and OEB use as the point estimator for $\theta_{\mathrm{FM},n}$.
ABCLOC uses the variation of
\begin{equation}
 V_{c,b}^\star=A_b^\star-cT_b^\star,\qquad c>0.
 \label{eqS:abcloc-basis}
\end{equation}
Let $\bar V_c^\star=B^{-1}\sum_{b=1}^B V_{c,b}^\star$, $I_{c,-}=\{b:V_{c,b}^\star\leq\bar V_c^\star\}$, and $I_{c,+}=\{b:V_{c,b}^\star\geq\bar V_c^\star\}$. Denote the numbers of replicates in these sets by $m_{c,-}$ and $m_{c,+}$, respectively.
In our implementation, the SDs below and above the bootstrap mean are calculated as
\begin{equation}
 s_{c,\pm}^2=
 \frac{1}{m_{c,\pm}}
 \sum_{b\in I_{c,\pm}}(V_{c,b}^\star-\bar V_c^\star)^2.
 \label{eqS:abcloc-spreads}
\end{equation}
With $z_p=\Phi^{-1}(p)$, where $\Phi$ is the standard normal distribution function, the interval is
\begin{equation}
 I_c=\left[
 \widehat C_{H,n}(P_n)-z_{1-\alpha/2}s_{c,+},\quad
 \widehat C_{H,n}(P_n)+z_{1-\alpha/2}s_{c,-}
 \right].
 \label{eqS:abcloc-interval}
\end{equation}
ABCLOC uses these two SDs and a standard normal quantile in place of the empirical quantiles used by NEB, OEB, TSB, and LSB.

Harrell selected $c=1.25$ by comparing absolute deviations of noncoverage in each tail from 0.025, summed over Somers' $D_{xy}=2\mathrm{AUC}-1$, calibration slope, and Brier score. The simulation used $n=200$ and 15 predictors. This was an empirical choice among the methods examined \citep{harrell2025}.
OEB instead follows from the estimation error for $\theta_{\mathrm{FM},n}$, but uses the same two performance statistics.

At $c=1$, $V_{1,b}^\star=O_b^\star$. Since $D_b^{\mathrm{OEB}}=A_n-V_{1,b}^\star$, the OEB replicate satisfies
\begin{equation}
 D_b^{\mathrm{OEB}}-\widehat C_{H,n}(P_n)
 =-(V_{1,b}^\star-\bar V_1^\star).
 \label{eqS:abcloc-oeb-exact}
\end{equation}
Thus, ABCLOC with $c=1$ replaces the OEB quantiles $q_{\alpha/2}\{D_b^{\mathrm{OEB}}\}$ and $q_{1-\alpha/2}\{D_b^{\mathrm{OEB}}\}$ with $\widehat C_{H,n}(P_n)-z_{1-\alpha/2}s_{1,+}$ and $\widehat C_{H,n}(P_n)+z_{1-\alpha/2}s_{1,-}$, respectively, where $q_u$ denotes the empirical $u$ quantile over bootstrap replicates.

\subsection{First-order coverage}\label{sup:abcloc-asymptotics}

Assume the expansions and center conditions in Section~S1.
We neglect Monte Carlo error by taking $B\to\infty$ before $n\to\infty$. Fitting failures are assumed to be asymptotically negligible, and the probability of successful interval construction is assumed to tend to one.

For a generic bootstrap sample, write
\[
 A^\star=\Psi(P_n^\star,P_n^\star),\qquad
 T^\star=\Psi(P_n^\star,P_n),\qquad
 A_n=\Psi(P_n,P_n).
\]
Let $V_c^\star=A^\star-cT^\star$ and recall that $\G_n^\star f=\sqrt n(P_n^\star-P_n)f$. Applying Equation~\eqref{eqS:local-expansion} with $(P,Q)=(P_n^\star,P_n^\star)$, $(P_n^\star,P_n)$, and $(P_n,P_n)$ gives, respectively,
\begin{align*}
 A^\star-\Psi(P_0,P_0)
   &=(P_n^\star-P_0)(\phi_{\mathrm{tr}}+\phi_{\mathrm{ev}})
     +o_{p^\star}(n^{-1/2}),\\
 T^\star-\Psi(P_0,P_0)
   &=(P_n^\star-P_0)\phi_{\mathrm{tr}}
     +(P_n-P_0)\phi_{\mathrm{ev}}
     +o_{p^\star}(n^{-1/2}),\\
 A_n-\Psi(P_0,P_0)
   &=(P_n-P_0)(\phi_{\mathrm{tr}}+\phi_{\mathrm{ev}})
     +o_p(n^{-1/2}),
\end{align*}
where the conditional orders hold in probability. Subtracting the third equation from each of the first two and multiplying by $\sqrt n$, we obtain
\begin{align}
 \sqrt n(A^\star-A_n)
   &=\G_n^\star(\phi_{\mathrm{tr}}+\phi_{\mathrm{ev}})
     +r_{A,n}^\star,\notag\\
 \sqrt n(T^\star-A_n)
   &=\G_n^\star\phi_{\mathrm{tr}}+r_{T,n}^\star,
 \label{eqS:abcloc-component-expansions}
\end{align}
where the remainders are assumed to satisfy
\begin{equation}
 \E_\star[(r_{A,n}^\star)^2+(r_{T,n}^\star)^2]=o_p(1).
 \label{eqS:abcloc-l2-remainders}
\end{equation}
Subtracting $c$ times the second line of Equation~\eqref{eqS:abcloc-component-expansions} from the first gives
\begin{equation}
 \sqrt n\{V_c^\star-(1-c)A_n\}
 =\G_n^\star\{\phi_{\mathrm{ev}}+(1-c)\phi_{\mathrm{tr}}\}
   +r_{A,n}^\star-cr_{T,n}^\star.
 \label{eqS:abcloc-uncentered-expansion}
\end{equation}
Since $\E_\star[\G_n^\star f]=0$, subtracting the conditional expectation yields
\begin{align}
 \sqrt n\{V_c^\star-\E_\star[V_c^\star]\}
 &=\G_n^\star\{\phi_{\mathrm{ev}}+(1-c)\phi_{\mathrm{tr}}\}\notag\\
 &\quad+r_{A,n}^\star-\E_\star[r_{A,n}^\star]
   -c\{r_{T,n}^\star-\E_\star[r_{T,n}^\star]\}\notag\\
 &=\G_n^\star\{\phi_{\mathrm{ev}}+(1-c)\phi_{\mathrm{tr}}\}
   +o_{p^\star}(1).
 \label{eqS:abcloc-expansion}
\end{align}

Let
\begin{equation}
 \sigma_c^2=\operatorname{Var}_{P_0}\{\phi_{\mathrm{ev}}+(1-c)\phi_{\mathrm{tr}}\},
 \qquad
 \sigma_{\mathrm{ev}}^2=\operatorname{Var}_{P_0}(\phi_{\mathrm{ev}}),
 \label{eqS:abcloc-variances}
\end{equation}
and assume both are positive. By the conditional bootstrap central limit theorem and Equation~\eqref{eqS:abcloc-expansion}, $\sqrt n\{V_c^\star-\E_\star[V_c^\star]\}\rightsquigarrow_\star N(0,\sigma_c^2)$ \citep{vanderVaart1998}. By symmetry of the limiting normal distribution and the conditional $L_2$ condition in Equation~\eqref{eqS:abcloc-l2-remainders}, the empirical SDs in Equation~\eqref{eqS:abcloc-spreads} satisfy
\begin{equation}
 \sqrt n\,s_{c,-}\xrightarrow{p}\sigma_c,
 \qquad
 \sqrt n\,s_{c,+}\xrightarrow{p}\sigma_c,
 \label{eqS:abcloc-sd-limits}
\end{equation}
as $B\to\infty$ followed by $n\to\infty$. For the final-model target, Section~S1 gives
\begin{equation}
 R_{\mathrm{FM},n}
 =\sqrt n\{C_{H,n}(P_n)-\theta_{\mathrm{FM},n}\}
 =\mathbb{G}_n\phi_{\mathrm{ev}}+o_p(1),
 \label{eqS:abcloc-target-error}
\end{equation}
so $R_{\mathrm{FM},n}/\sigma_{\mathrm{ev}}\rightsquigarrow N(0,1)$. The interval in Equation~\eqref{eqS:abcloc-interval} covers $\theta_{\mathrm{FM},n}$ if and only if
\begin{equation}
 -z_{1-\alpha/2}\sqrt n\,s_{c,-}
 \leq\sqrt n\{\widehat C_{H,n}(P_n)-\theta_{\mathrm{FM},n}\}
 \leq z_{1-\alpha/2}\sqrt n\,s_{c,+}.
 \label{eqS:abcloc-coverage-event}
\end{equation}
Dividing by $\sigma_{\mathrm{ev}}>0$, we obtain the equivalent event
\[
 -z_{1-\alpha/2}\frac{\sqrt n\,s_{c,-}}{\sigma_{\mathrm{ev}}}
 \leq
 \frac{\sqrt n\{\widehat C_{H,n}(P_n)-\theta_{\mathrm{FM},n}\}}
      {\sigma_{\mathrm{ev}}}
 \leq
 z_{1-\alpha/2}\frac{\sqrt n\,s_{c,+}}{\sigma_{\mathrm{ev}}}.
\]
For each fixed $n$, letting $B\to\infty$ replaces the middle term by $R_{\mathrm{FM},n}/\sigma_{\mathrm{ev}}$. As $n\to\infty$, this standardized error converges in distribution to $N(0,1)$, while the lower and upper bounds converge in probability to $-z_{1-\alpha/2}\sigma_c/\sigma_{\mathrm{ev}}$ and $z_{1-\alpha/2}\sigma_c/\sigma_{\mathrm{ev}}$, respectively. Slutsky's lemma therefore gives
\begin{align}
 \lim_{n\to\infty}\lim_{B\to\infty}
 \Pr\{\theta_{\mathrm{FM},n}\in I_c\}
 &=
 \Phi\!\left(z_{1-\alpha/2}\frac{\sigma_c}{\sigma_{\mathrm{ev}}}\right)
 -
 \Phi\!\left(-z_{1-\alpha/2}\frac{\sigma_c}{\sigma_{\mathrm{ev}}}\right)
 \notag\\
 &=2\Phi\!\left(z_{1-\alpha/2}\frac{\sigma_c}{\sigma_{\mathrm{ev}}}\right)-1,
 \label{eqS:abcloc-general-coverage}
\end{align}
where $\Phi$ is the standard normal distribution function. Thus, $c=1$ yields asymptotic coverage $1-\alpha$. If $\phi_{\mathrm{tr}}=0$ $P_0$-almost surely, the same holds for every fixed $c>0$.

In particular, at $c=1$,
\begin{align}
 \sqrt n\{V_1^\star-\E_\star[V_1^\star]\}
   &=\G_n^\star\phi_{\mathrm{ev}}+o_{p^\star}(1),\notag\\
 \sqrt n\{D_b^{\mathrm{OEB}}-C_{H,n}(P_n)\}
   &=-\G_n^\star\phi_{\mathrm{ev}}+o_{p^\star}(1).
 \label{eqS:abcloc-sign}
\end{align}
The signs differ, but the centered Gaussian limit is unchanged by reflection. Hence, ABCLOC with $c=1$ and OEB have the same first-order limits for their interval endpoints.

\begin{figure}[p]
\centering
\includegraphics[
  width=\textwidth,
  height=0.65\textheight,
  keepaspectratio
]{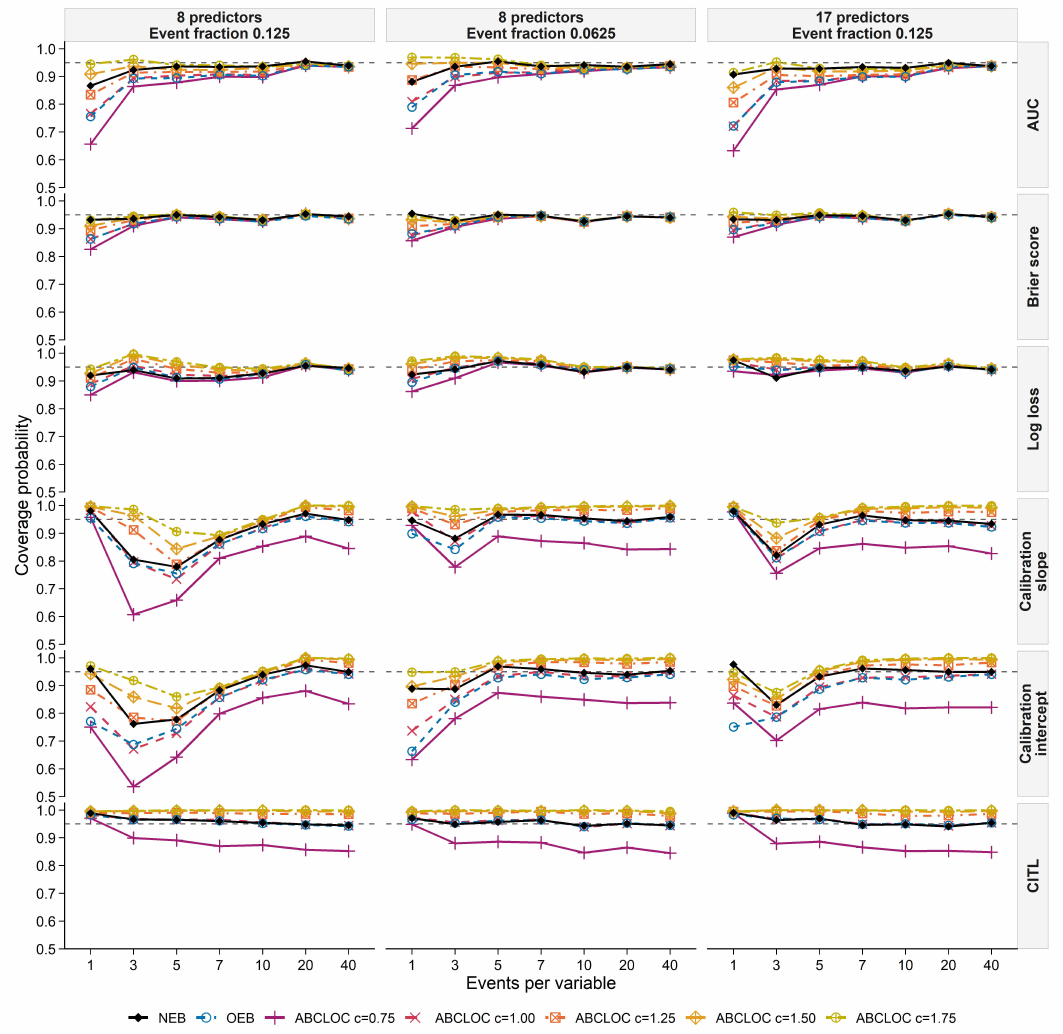}
\vspace{2pt}
\caption{Coverage for $\theta_{\mathrm{FM},n}$ using asymmetric bootstrap confidence limits for overfitting-corrected model performance metrics (ABCLOC), nested error bootstrap (NEB), and optimism error bootstrap (OEB) in Simulation~1 (GUSTO-I-based settings). Empirical coverage of 95\% confidence intervals (CIs) is shown across events per variable (EPV) 1, 3, 5, 7, 10, 20, and 40. Coverage was evaluated among replicates in which a CI was successfully constructed and the target value was available for each method. Columns represent the settings with 8 predictors and event fraction 0.125, 8 predictors and event fraction 0.0625, and 17 predictors and event fraction 0.125. Rows show area under the receiver operating characteristic curve (AUC), Brier score, log loss, calibration slope, calibration intercept, and calibration-in-the-large (CITL). Curves show ABCLOC with $c=0.75,1.00,1.25,1.50,$ and $1.75$, NEB, and OEB; the dashed horizontal line denotes 0.95.}
\label{figS:abcloc}
\end{figure}

As shown in Section~S1, $\phi_{\mathrm{tr}}=0$ for log loss under unpenalized logistic maximum likelihood and for Brier score under correct logistic specification. Under the conditions above, their limiting coverage therefore does not depend on $c$. For AUC, the additional conditions in Section~S3 are required, and its finite-sample behavior is assessed by simulation. Supplementary Figure~\ref{figS:abcloc} compares ABCLOC with NEB and OEB in Simulation~1 (GUSTO-I-based settings). At moderate and higher EPV, ABCLOC with $c=1.25$ gave coverage broadly similar to NEB and OEB for AUC, Brier score, and log loss.

For the calibration measures, $c$ directly affects the limiting coverage. Under the same regularity conditions, the identities in Appendix~A of the main article for unpenalized logistic regression with an intercept give
\begin{equation}
 \Psi_{\mathrm{cal}}(P,P)=c_0,\qquad
 \phi_{\mathrm{tr}}^{\mathrm{cal}}=-\phi_{\mathrm{ev}}^{\mathrm{cal}}
 \quad P_0\text{-almost surely},
 \label{eqS:abcloc-calibration-identity}
\end{equation}
where $c_0=1$ for calibration slope and $c_0=0$ for calibration intercept and CITL.
Combining Equations~\eqref{eqS:abcloc-expansion} and~\eqref{eqS:abcloc-calibration-identity} gives
\begin{equation}
 \sqrt n\{V_c^\star-\E_\star[V_c^\star]\}
 =c\G_n^\star\phi_{\mathrm{ev}}^{\mathrm{cal}}+o_{p^\star}(1),
 \qquad \sigma_c=c\sigma_{\mathrm{ev}}.
 \label{eqS:abcloc-calibration-root}
\end{equation}
Hence, for these calibration measures,
\begin{equation}
 \lim_{n\to\infty}\lim_{B\to\infty}
 \Pr\{\theta_{\mathrm{FM},n}^{\mathrm{cal}}\in I_c\}
 =2\Phi(cz_{1-\alpha/2})-1.
 \label{eqS:abcloc-calibration-coverage}
\end{equation}
Among positive values of $c$, nominal coverage is attained uniquely at $c=1$.
For nominal 95\% intervals, the limits are 
\[
\begin{array}{c|ccccc}
c&0.75&1.00&1.25&1.50&1.75\\[2pt]
\text{Limiting coverage}&0.8584&0.9500&0.9857&0.9967&0.9994.
\end{array}
\]
Indeed, when $A_b^\star=c_0$ for every retained bootstrap sample, $V_{c,b}^\star-\bar V_c^\star=c(V_{1,b}^\star-\bar V_1^\star)$ exactly, so both SDs are multiplied by $c$. Thus, $c>1$ gives asymptotic overcoverage, whereas $0<c<1$ gives undercoverage. In Supplementary Figure~\ref{figS:abcloc}, coverage for the calibration measures at higher EPV increased with $c$, consistent with Equation~\eqref{eqS:abcloc-calibration-coverage}. With $c=1$, coverage generally followed that of OEB. Differences at low EPV varied by measure and setting and did not show a consistent advantage.

For inference on $\theta_{\mathrm{FM},n}$ under the conditions above, we therefore recommend $c=1$ when ABCLOC is used across performance measures. This choice preserves the first-order variation required for calibration as well as for the other measures. In some low-EPV scenarios, such as those with EPV = 1, ABCLOC with $c=1$ showed slightly higher coverage than OEB. However, little difference in coverage was observed in the practically relevant settings with EPV $\geq10$. Further investigation is needed to determine whether using separate lower- and upper-tail SDs, as in ABCLOC, can improve coverage for other methods.

\clearpage
\renewcommand{\refname}{References}
\bibliographystyle{oup-abbrvnat}
\bibliography{references}

\clearpage

\clearpage

\clearpage

\clearpage
\begin{figure}[p]
\centering
\makebox[\linewidth][c]{%
  \includegraphics[
    width=\linewidth,
    height=0.60\textheight,
    keepaspectratio
  ]{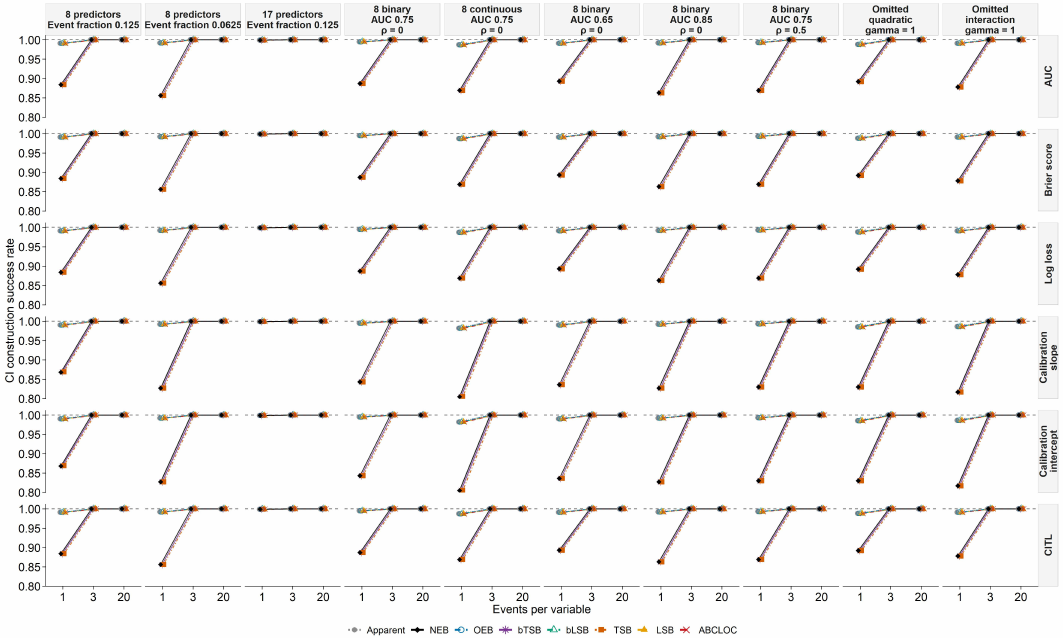}%
}
\vspace{-4pt}
\caption{Success rates for constructing confidence intervals (CIs) in Simulations~1 and~2. Empirical proportions of simulation replicates with successful construction of 95\% CIs are shown at events per variable (EPV) 1, 3, and 20. The first three columns represent the GUSTO-I-based settings of Simulation~1 with 8 predictors and event fraction 0.125, 8 predictors and event fraction 0.0625, and 17 predictors and event fraction 0.125. The remaining columns represent the synthetic settings of Simulation~2: the baseline setting with 8 independent Bernoulli$(0.5)$ predictors and population area under the receiver operating characteristic curve (AUC) 0.75, the setting with 8 continuous predictors, the settings with population AUC 0.65 and 0.85, the setting with a Pearson correlation of 0.5 among the binary predictors, and the settings with an omitted quadratic or interaction term ($\gamma=1$). Rows show AUC, Brier score, log loss, calibration slope, calibration intercept, and calibration-in-the-large (CITL). Curves show Apparent, nested error bootstrap (NEB), optimism error bootstrap (OEB), basic two-stage bootstrap (bTSB), basic location-shifted bootstrap (bLSB), two-stage bootstrap (TSB), location-shifted bootstrap (LSB), and asymmetric bootstrap confidence limits for overfitting-corrected model performance metrics (ABCLOC) with $c=1$; the dashed horizontal line denotes 1.00. Results at EPV 5, 7, 10, and 40 are omitted because all methods achieved 100\% CI construction success across the settings and performance measures shown.}
\label{figS:ci-success}
\end{figure}